\documentclass[amssymb,nobibnotes,nofootinbib,aps,preprintnumbers,prd]{revtex4}

 \usepackage{here}

\usepackage[dvipdfmx]{graphicx}
\usepackage{amssymb,amsfonts,amsmath,bm,color,multirow,cases,empheq,hyperref}
\usepackage{mathrsfs}
\definecolor{darkergreen}{rgb}{0,0.5,0}

\usepackage{enumerate}
\usepackage{ulem}
\usepackage{afterpage}
\hypersetup{%
 setpagesize=false,
 bookmarksnumbered=true,%
 bookmarksopen=true,%
 colorlinks=true,%
 linkcolor=blue,%
 citecolor=red}
\newcommand{\fr}[1]{\frac{1}{#1}}

\newcommand{\cC}{{\mathcal C}}

\newcommand{\cD}{{\mathcal D}}

\newcommand{\nonum}{\nonumber\\ }

\newcommand{\cout}[1]{}

\newcommand{\arrayL}[1]{\left(\begin{array}{#1}}
\newcommand{\arrayR}{\end{array}\right)}
\newcommand{\arrayLb}[1]{\left[\begin{array}{#1}}
\newcommand{\arrayRb}{\end{array}\right]}

\begin{document}
\title{Nonuniqueness of capped black holes: large and small bubbles
}

\author{Ryotaku Suzuki}
\email{sryotaku@toyota-ti.ac.jp}
\author{Shinya Tomizawa}
\email{tomizawa@toyota-ti.ac.jp}
\affiliation{\vspace{3mm}Mathematical Physics Laboratory, Toyota Technological Institute\vspace{2mm}\\Hisakata 2-12-1, Tempaku-ku, Nagoya, Japan 468-8511\vspace{3mm}}

\begin{abstract}

We present a new non-BPS solution describing an asymptotically flat, stationary, bi-axisymmetric capped black hole  in the bosonic sector of five-dimensional minimal supergravity.
This solution describes a spherical black hole, while the exterior region of the horizon exhibits a non-trivial topology of $[{\mathbb R}^4 \# {\mathbb C}{\mathbb P}^2] \setminus {\mathbb B}^4$ on a timeslice.
This solution extends our previously constructed three-parameter solution to a more general four-parameter solution.
To derive this solution, we utilize a combination of the Ehlers and Harrison transformations and then impose appropriate  boundary conditions on the solution's parameters.
It can be shown that the resultant solution is free from curvature, conical, Dirac-Misner string and orbifold singularities, as well as closed timelike curves on and outside the horizon. 
Characterized by four independent conserved charges---mass, two angular momenta, and electric charge---this solution reveals two distinct branches: a small bubble branch and a large bubble branch, distinguished by non-conserved local quantities such as magnetic flux or magnetic potential.
This shows the non-uniqueness for spherical black holes, even among capped black holes.
For equivalent sets of conserved charges, we find that the large/small bubble branch can have  larger/smaller entropy than the Cveti\v{c}-Youm black hole.
\end{abstract}

\date{\today}
\preprint{TTI-MATHPHYS-32}

\maketitle

\section{Introduction}

The topological censorship theorem imposes significant constraints on the allowed topology of the domain of outer communication (DOC)~\cite{Friedman:1993ty}. In four dimensions, only a trivial topology represented by ${\rm DOC} \cap \Sigma = \mathbb{R}^3 \setminus \mathbb{B}^3$ ($\Sigma$: timeslice) is permissible. However, in higher dimensions, non-trivial topologies that may have homology groups of higher rank are allowed. 
From this theorem, Ref.~\cite{Hollands:2010qy} demonstrated that in a five-dimensional stationary and axisymmetric spacetime, ${\rm DOC} \cap \Sigma$ can admit the non-trivial topology of $[\mathbb{R}^4 \# n (\pm \mathbb{C}{\mathbb P}^2) \# m(S^2 \times S^2)] \setminus \mathbb{B}^4$.
Despite these possibilities, in higher dimensions for static asymptotically flat spacetimes, no black hole solution with such nontrivial DOC exists due to the uniqueness theorems~\cite{Gibbons:2002bh,Gibbons:2002av}. 
These theorems assert that the higher-dimensional Schwarzschild and Reissner-Nordstr\"om solutions~\cite{Tangherlini:1963bw} are the only vacuum and charged black hole solutions, respectively. 
The uniqueness theorem for charged rotating black holes in the bosonic sector of five-dimensional minimal supergravity~\cite{Tomizawa:2009ua} states that under the assumptions of two commuting axial isometries and spherical topology of horizon cross-sections, an asymptotically flat, stationary charged rotating black hole with a non-extremal horizon is uniquely characterized by its mass, charge, and two independent angular momenta and is therefore described by the five-dimensional Cveti\v{c}-Youm solution~\cite{Cvetic:1996xz}.
However, unlike the uniqueness theorem in four dimensions, this theorem assumes the additional condition that the intersection of a black hole's exterior region with the timeslice $\Sigma$ has a trivial topology, $\mathbb{R}^4 \setminus \mathbb{B}^4$. Thus, it is still possible that such a solution with DOC of nontrivial topologies exists, and if it exists, it cannot be static but must belong to a class of stationary solutions.

\medskip
Kunduri and Lucietti have constructed a BPS (Bogomol'nyi-Prasad-Sommerfield) black hole solution within the bosonic sector of five-dimensional minimal supergravity, which features  the DOC of the topology $[{\mathbb R}^4 \# (S^2\times S^2) ]\setminus {\mathbb B}^4$~\cite{Kunduri:2014iga}.
In our previous work, we developed the first non-BPS exact solution representing an asymptotically flat, stationary spherical black hole with a nontrivial topology of $[\mathbb{R}^4 \# \mathbb{CP}^2] \setminus \mathbb{B}^4$ in five-dimensional minimal supergravity.
This solution was derived using a methodology similar to that employed for the charged dipole black ring previously constructed in our work~\cite{Suzuki:2024coe}. 
Initially, we utilized the inverse scattering method (ISM) to construct a vacuum black hole possessing a Dirac-Misner string singularity. Following this, we applied the electric Harrison transformation to this vacuum solution, resulting in a charged rotating solution characterized by an $S^3$ horizon topology, yet still retaining the Dirac-Misner string singularity.
To ensure the regularity of the solution, we finely tuned the solution's parameters to eliminate the Dirac-Misner string singularity. 
The final solution has independent three parameters and exhibits regularity, with the absence of curvature, conical, Dirac-Misner string, and orbifold singularities, both inside and outside the horizon, as well as the absence of closed timelike curves (CTCs). 
It describes a charged rotating black hole capped by a disk-shaped bubble, hence we called it ``capped black hole".

\medskip
In this paper, using the Ehlers-Harrison transformations, previously applied to the charged rating black ring with all independent conserved charges~\cite{Suzuki:2024vzq},
we wish to generalize the three-parameter family of the capped black hole solution, derived in our previous works~\cite{Suzuki:2023nqf,Suzuki:2024phv}, to a four-parameter family. 
The Ehlers and Harrison transformations add an angular momentum and an electric charge to a five-dimensional vacuum solution, respectively, preserving asymptotic flatness. 
In our previous work, we developed the construction of a vacuum black ring solution possessing a Dirac-Misner string singularity through the Ehlers transformation, demonstrating that a doubly rotating black ring (Pomeransky-Sen'kov solution~\cite{Pomeransky:2006bd}) can be derived once the Dirac-Misner string singularity is removed. 
Subsequently, in Ref.~\cite{Suzuki:2024vzq}, we applied the Harrison transformation to this vacuum solution, generating a corresponding charged solution within five-dimensional minimal supergravity. 
By setting appropriate boundary conditions on the parameters, we successfully eliminated all singularities—including those related to curvature, conical features, and Dirac-Misner strings—across the rotational axes and the horizon. 
Using this methodology, we can construct a capped black hole, distinguishing it by replacing the $S^1\times S^2$ topology with $S^3$.
We demonstrate that the new solution manifests discrete non-uniqueness in terms of conserved
 charges, presenting two distinct branches that share the same mass, angular momenta, and electric charges as the Cveti\v{c}-Youm black hole. These branches are distinguished by the size of their bubbles adjacent to the horizon and are accordingly termed {\it small bubble} and {\it large bubble} branches. Our analysis indicates that within one parameter region, the large bubble branch exhibits greater entropy than both the small bubble branch and the Cveti\v{c}-Youm black hole, making it thermodynamically more favored. Conversely, in another parameter region, the Cveti\v{c}-Youm black hole displays higher entropy than both the large and small bubble branches.

\medskip
The rest of the paper is organized as follows.
In section~\ref{sec:review}, 
we present  the metric and gauge potential for the new non-BPS solution of the capped black hole in  the C-metric coordinates.
In section~\ref{sec:bdry}, we impose the boundary conditions for the absence of curvature, conical, orbifold, Dirac-Misner string singularities and closed timelike curves on the solution's parameters.
In section~\ref{sec:phase}, we investigate the phase of the physical solution. 
Finally, we summarize the results in section~\ref{sec:sum}.

\section{Solution generation\label{sec:review}}

We consider asymptotically flat, stationary, and bi-axisymmetric solutions in the bosonic sector of five-dimensional minimal ungauged supergravity (Einstein-Maxwell-Chern-Simons theory), whose the action is expressed as follows
\begin{eqnarray}
S=\frac{1}{16 \pi G_5}  \left[ 
        \int d^5x \sqrt{-g}\left(R-\frac{1}{4}F^2\right) 
       -\frac{1}{3\sqrt{3}} \int F\wedge F\wedge A 
  \right] \,, 
\label{action} 
\end{eqnarray} 
where $F=dA$. The field equation comprises of the Einstein equation 
\begin{eqnarray}
 R_{\mu \nu } -\frac{1}{2} R g_{\mu \nu } 
 = \frac{1}{2} \left( F_{\mu \lambda } F_\nu^{ ~ \lambda } 
  - \frac{1}{4} g_{\mu \nu } F_{\rho \sigma } F^{\rho \sigma } \right) \,, 
 \label{Eineq}
\end{eqnarray}
and the Maxwell equation with a Chern-Simons term
\begin{eqnarray}
 d\star F+\frac{1}{\sqrt{3}}F\wedge F=0 \,. 
\label{Maxeq}
\end{eqnarray}
Assuming that there is one timelike Killing vector $\xi_0 = \partial/\partial t$ and one spacelike axial Killing vector $\xi_1 = \partial/\partial \psi$, this theory simplifies to the $G_{2(2)}/SL(2,{\mathbb{R}}) \times SL(2,{\mathbb{R}})$ non-linear sigma models coupled with three-dimensional gravity~\cite{Mizoguchi:1998wv,Mizoguchi:1999fu}. 
By further assuming the existence of a third spacelike axial Killing vector $\xi_2 = \partial/\partial \phi$ commuting with the other two, the theory reduces to the $G_{2(2)}/[SL(2,\mathbb{R})\times SL(2,\mathbb{R})]$ nonlinear sigma model.
Because the theory is invariant under the $G_{2(2)}$-transformation, one can produce a new solution from a known solution by applying a certain transformation in the $G_{2(2)}$-group.
Below, we present the solution derived through two different $G_{2(2)}$-transformation: the Ehlers transformation~(Eq.~(2.45) in Ref.~\cite{Giusto:2007fx}, or Eq.~(32) in Ref.~\cite{Suzuki:2024eoz} for precise expression), and Harrison transformation~(Eq.~(119) in Ref.~\cite{Bouchareb:2007ax}), where the former adds the angular momenta to the five-dimensional asymptotically flat vacuum solution and the latter the electric charge.

\medskip
We aim to obtain a generalized version of the capped black hole obtained in Ref.~\cite{Suzuki:2023nqf,Suzuki:2024phv}.
To this end, we prepare the charged solution by successively applying the Ehlers and Harrison transformation to the vacuum seed solution produced by the inverse scattering method in Ref.~\cite{Chen:2008fa}.
This charged solution was first constructed in our previous work~\cite{Suzuki:2024vzq}, to derive the charged black ring solution.
We show that the same solution also contains the capped black hole under the different boundary condition.
Below, we start with revisiting the charged solution obtained in Ref.~\cite{Suzuki:2024vzq}, whose metric and gauge potential are given by
\begin{align}
&ds^2 = - \frac{H(y,x)}{D^2H(x,y)} (dt+\Omega'_\psi(x,y) d\psi + \Omega'_\phi(x,y)d\phi)^2+\frac{D}{H(y,x)}\left[F(y,x)d\psi^2-2J(x,y)d\psi d\phi-F(x,y)d\phi^2\right]\nonum
&\quad + \frac{\ell^2 D H(x,y)}{4(1-\gamma)^3(1-\nu^2)(1-a^2)\Delta (x-y)^2}\left(\frac{dx^2}{G(x)}-\frac{dy^2}{G(y)}\right),
\label{eq:metricsol}
\end{align}
and
\begin{align}\label{eq:gaugesol}
&A = \frac{\sqrt{3}cs}{DH(x,y)} \left[ (H(x,y)-H(y,x))dt \right.\nonum
&\hspace{2cm} \left.- (c H(y,x) \Omega_\psi(x,y)-s H(x,y) \Omega_\phi(y,x))d\psi
- (c H(y,x) \Omega_\phi(x,y)-s H(x,y) \Omega_\psi(y,x))d\phi\right],
\end{align}
with
\begin{align}
D &:= \frac{c^2 H(x,y)-s^2 H(y,x)}{H(x,y)},\\
\Omega_\psi'(x,y) &:= c^3 \Omega_\psi(x,y)-s^3 \Omega_\phi(y,x),\\
\Omega_\phi'(x,y) &:= c^3\Omega_\phi(x,y)-s^3\Omega_\psi(y,x),
\end{align}
where the parameter $\alpha$ induced by the Harrison transformation appears as $(c,s):=(\cosh\alpha,\sinh\alpha)$.
The metric functions become
\begin{align}
\label{eq:defGx}
G(u) &= (1-u^2)(1+\nu u),\\
 H (x,y) &=(1+y)^2 \biggr[ \nu  d_1 f_7^2 (1-\nu )  \left(1-x^2\right) \left(2 (\gamma -1) \nu +c_3\right)+ 2 c_3 c_1^2 f_6^2 (\gamma -\nu) (\nu  x+1)^2 \nonum
&\quad  -(\gamma -\nu ) (1-\nu )  \left(c_1^2 c_3 f_6^2-2 (1-\gamma ) \nu  \left(c_1-b (1-\nu   )\right){}^2 f_8^2\right) (1-x) (1+x \nu )\biggr]\nonum
&  -(1+y)  \biggr[ d_5 g_6 (1-x)^2  +d_6 g_7 \left(1-x^2\right)    +2(1-\nu)^{-1} d_6  g_8(1+x)  (1+x \nu )  \biggr] \nonum
& + 4 \left(1-a^2\right) (1-x) (1-\gamma )^3 (1-\nu )^4 \Delta-4 (1+x) (1-\gamma ) (1-\nu )^2 (1+\nu ) d_1 f_2^2+2 \left(1-x^2\right) (1-\gamma ) (1-\nu
   )^2 c_2^2 f_5^2,  \label{eq:Hxy}\\
F(x,y)&=\frac{\ell^2 \tilde{v}_0^2(1-\gamma)}{(\gamma^2-\nu^2) (x-y)^2}\biggr[
4 (1+y \nu ) G(x) \left\{ \left(1-a^2\right)^2 (-1+y) (1-\gamma )^3 (1-\nu )^3 \Delta^2-(1+y) d_1^2 f_1^2 f_2^2\right\}\nonum
&+4 (1+x) (1+x \nu )  G(y) \left((1-a b) (\gamma -\nu ) (1+\nu ) c_1 -(1-\nu ) c_2 \right){}^2 g_{5}^2\nonum
&+ \nu^{-1} (1-\nu )^3 (\gamma -\nu ) \biggr\{\left(y^2-1\right) G(x) c_3^2 f_3^2 f_4^2+\left(1-x^2\right) G(y) d_3^2 g_2^2\biggr\}\nonum
& + G(x) G(y) \biggr\{\frac{x (\gamma -\nu ) (c_1 c_3f_3 f_6-b d_1f_1f_7)^2 }{1-\gamma }-(a-b)^2 y (1-\gamma ) (\gamma -\nu ) c_2^2
   f_5^2 f_8^2+\frac{(1-a^2) d_4 \Delta g_4}{\nu }\biggr\}   
\biggr],\\
J(x,y)& =  \frac{ \ell^2 \tilde{v}_0^2 (1-\gamma)(1+x)(1+y)}{(\gamma^2-\nu^2)(x-y)}
\biggr[4 d_1 f_1 f_2 g_{5} \biggr\{(1-a b) (\gamma -\nu ) (1+\nu ) c_1 -(1-\nu ) c_2 \biggr\} (1+x \nu ) (1+y \nu ) \nonum
& - (1-x) (1-y) (\gamma -\nu ) \biggr\{(a-b) (1+x \nu ) (1+y \nu ) c_2 f_5 f_8 \left(c_1 c_3 f_3 f_6 -b d_1 f_1 f_7 \right) 
+(1-\nu )^3 c_3   d_3 f_3 f_4 g_2\biggr\}
\biggr],\\
\Omega_\psi (x,y)&= \frac{\ell \tilde{v}_0 (1+y)}{  H (y,x)}\biggr[
\nu^{-1}c_2 f_5 \left(c_1 c_3 f_4 f_{16}-b d_1 f_2 f_{13}\right)  (1-x) (1-\nu ) (1+x \nu ) (1+y \nu ) \nonum
&+ (1+y \nu ) \left\{(1+x) (1+x \nu ) c_1 c_3 f_2 f_9 f_{10}-\frac{1}{2} b \left(1-x^2\right) (\gamma -\nu ) (1-\nu ) c_3 f_4 f_{10} f_{12}\right\}\nonum
&-2 (1+x) (1-\nu ) (1+x \nu ) d_1 f_2 f_{11}  f_{15} \left(c_3-(1-a b) (1-\gamma ) (1+\nu ) \right)\nonum
&+\nu^{-1} c_3 \left(d_1-d_2\right) f_4 g_3(1-x) (1-\nu )^2 (1+x \nu ) +c_3 d_2 f_4 g_{10} (1-\nu )^2\left(1-x^2\right) 
 \biggr],\\
\Omega_\phi(x,y) &= \frac{\ell \tilde{v}_0 (1+x)}{ H (y,x)}\biggr[
b (1+x) (1+y) (1+y \nu ) d_1 d_2 f_1 f_{13} f_{14}\nonum
& +\frac{(1+x) \left( 1-y^2\right) (1-\nu ) \nu  c_3 \left(b (1+\nu ) d_1-2 (a-b) (1-\gamma )^2 (1-\nu ) \nu \right) f_3 g_{11}}{1+\nu }\nonum
&+\frac{(y-1) (1-\gamma ) (1-\nu ) \nu  (1+y \nu + x (1+y (4-3 \nu ))) c_3 f_1 f_3 f_{10}}{1+\nu }\nonum
&+\frac{2 (a-b) (1-\gamma )^2 \left((y-1) (x+y) (1-\nu )^3 \nu  c_3 f_3 g_{9}+2 (1+x \nu ) (1+y \nu )^2 d_1 f_1 g_{1}\right)}{1+\nu }
\biggr],
\end{align}
where 
\begin{align}\label{eq:def-cfs}
\begin{split}
&\tilde{v}_0 := \frac{v_0}{\sqrt{\Delta}} := \sqrt{\frac{2(\gamma^2-\nu^2)}{(1-a^2)(1-\gamma)\Delta}},\\
&c _1 :=(1-\gamma) a+(\gamma-\nu) b ,\\
&c _2 := 2 a (1-\gamma ) \nu +b (\gamma -\nu ) (1+\nu ),\\
&c _3 := 2   (1-\gamma ) \nu +b^2 (\gamma -\nu ) (1+\nu ),\\
&d_1 := (\nu +1) c _1^2-(1-\gamma ) (1-\nu )^2,\\
&d_2:= b (\nu +1) c _1 (\gamma -\nu )+2\nu (1-\gamma )   (1-\nu ) ,\\
&  d_3 := \left(a^2-1\right) b (\gamma -1) (\nu +1)-a c _3,\\
&d_4 := b^2 (\gamma -\nu )   \left[(\nu +1)^2 c _1^2 \left(-3 (1-\gamma ) \nu -\nu ^2+1\right)-(1-\gamma ) (1-\nu )^4 (2 \nu +1)\right]\\
&\quad +(1-\gamma   ) \left[\left((1-\nu ) c _2-2 \nu ^2 c _1\right)^2-4 \nu ^2 c _1^2 \left(-\gamma  (\nu +2)+3 \nu
   ^2+1\right)\right],\\
& d_5 := (1-\gamma ) (1-\nu )^3 [(\gamma -3 \nu ) \left(b^2 (\nu -1) (\gamma -\nu )-c_1^2\right)
-2 b c_1 (3 \nu -1) (\gamma -\nu )],\\
& d_6 := c_1(1-\gamma ) (1-\nu) \left(1-\nu ^2\right) \left(c_2-(1-\gamma ) (a-b) (\gamma -\nu )\right),\\
&\Delta := f_8^2 - \frac{d_1(f_1^2-f_8^2)}{(1-a^2)(1-\gamma)(1-\nu^2)}+\frac{c_3 (f_3^2-f_8^2)}{(1-a^2)(1-\gamma)(1+\nu)},
   \end{split}
\end{align}
and the metric depends on the parameter $\beta$ induced by the Ehlers transformation through coefficients $\{ f_i\}_{i=1,\dots 16}$ and $\{ g_i\}_{i=1,\dots 11}$, which are quadratic and quartic polynomials of $\beta$ presented in Appendix~\ref{sec:fi-gi}, respectively.

\medskip
We assume that the coordinates $(t,\psi,\phi,x,y)$ run in the ranges,
\begin{eqnarray}
-\infty < t <\infty, \quad 0\leq \psi\le 2\pi, \quad 0\le \phi \leq 2\pi, \label{eq:tphipsirange}
\end{eqnarray}
and
\begin{align}
 -1 \leq x \leq 1,\quad -1/\nu \leq y \leq -1. \label{eq:xyrange}
\end{align}
The solution is described by seven parameters $(\ell,\nu,\gamma,a,b,\beta,\alpha)$, among which the parameters $\ell,\nu,\gamma$, are assumed to lie within the following range 
\begin{align}\label{eq:rangenugam}
\ell>0,\quad  0 < \nu < \gamma < 1,
\end{align}
while the ranges for $a,b,\beta,\alpha$ are not specified explicitly at this point. 
Moreover $\tilde{v}_0^2$ must be real so that the metric functions take real values, 
which leads to 
\begin{align}\label{eq:necessary-1}
(1-a^2)\Delta >0.
\end{align}

\medskip

\section{Boundary conditions}\label{sec:bdry}

In order that the solution~(\ref{eq:metricsol}) and (\ref{eq:gaugesol}) describes a capped black hole of physical interest, we must impose  the appropriate boundary conditions at
\begin{itemize}
\item[(i)] infinity $\partial \Sigma_\infty 
= \{(x,y)|x\to y\to -1 \}$,
\item[(ii)] the $\phi$-rotational axis $\partial \Sigma_\phi=\{(x,y)|x=-1,-1/\nu<y<-1\}$,
\item[(iii)] the $\psi$-rotational axis $\partial \Sigma_\psi=\{(x,y)|-1<x<1,y=-1\}$,
\item[(iv)] the inner axis (bubble) $\partial \Sigma_{\rm in}=\{(x,y)|x=1,-1/\nu<y<-1\}$,
\item[(vi)] the center at $(x,y)=(1,-1)$, 
\item[(v)] the horizon $\partial \Sigma_{\cal H}=\{(x,y)|-1<x<1,y=-1/\nu\}$.
\end{itemize}
The boundary condition at infinity (i) is imposed so that the spacetime is asymptotically flat.
We require that on the axes (ii), (iii) and (iv), there appear no Dirac-Misner strings, and orbifold singularities at isolated points must be eliminated. 
We also demand that the horizon (v) describes a smooth null surface whose spatial cross section has a topology of $S^3$. 
At these boundaries, the spacetime is required to allow neither CTCs nor (conical and curvature) singularities.
Since the Kretchman invariant can be written as $R_{\mu\nu\rho\sigma}R^{\mu\nu\rho\sigma} \sim H(x,y)^{-6}D^{-6}$, 
the metric~(\ref{eq:metricsol}) exhibits the curvature singularity if there exist the surface $H(x,y)=0$ or $D=0$ on and outside the horizon. 
Henceforth, we must choose the parameters suitably such that  $H(x,y)\neq 0$ and $D\neq 0$ in the region~(\ref{eq:xyrange}) including the boundary. 
Since we observe from Eqs.~(\ref{eq:rangenugam}) and (\ref{eq:necessary-1}) that at infinity, 
\begin{align}
&H(x=-1,y=-1) = 8(1-\gamma)^3(1-\nu)^4(1-a^2)\Delta >0, \quad D(x=-1,y=-1)=1>0,
\end{align}
 the requirements $H(x,y)\neq 0$ and $D\neq 0$ on and outside the horizon can be replaced with $H(x,y)>0$ and $D>0$,

\subsection{Asymptotic infinity ($x\to y \to -1$)}

The boundary $\partial \Sigma_\infty 
= \{(x,y)|x\to y\to -1 \}$ corresponds to the asymptotic infinity.
This can be seen in  the $(r,\theta)$ coordinates introduced by
\begin{align}
 x = -1 + \frac{4 \ell^2 (1-\nu) \cos^2\theta }{r^2},\quad y = -1 - \frac{4\ell^2 (1-\nu) \sin^2\theta}{r^2},\label{eq:asym-xy}
\end{align}
in which the metric~(\ref{eq:metricsol}) asymptotes to the five-dimensional Minkowski metric at the limit $r\to\infty$
\begin{align}\label{eq:asym-charges}
ds^2 &\simeq -\left(1-\frac{8G_5 M}{3\pi r^2}\right)dt^2-\frac{8G_5 J_\psi \sin^2\theta}{\pi r^2}dt d\psi-\frac{8G_5 J_\phi \cos^2\theta}{\pi r^2}dt d\phi\nonum
& +dr^2 + r^2\sin^2\theta d\psi^2+r^2 \cos^2\theta d\phi^2+r^2d\theta^2,
\end{align}
where $M,J_\psi,J_\phi$ are the ADM mass and angular momenta which are expressed as
\begin{align}
 M &= (c^2+s^2)   \frac{3  \pi \ell^2 \tilde{v}_0^2  \left(d_1-(1-\nu ) c_3\right)}{8 G_5  (1-\gamma ) (1+\nu )(\gamma-\nu)} \biggr[
f_8^2+\frac{(1+\nu ) c_2^2 \left(f_8^2-f_5^2\right)}{(1-\nu ) (\gamma +\nu ) \left(d_1-(1-\nu )   c_3\right)}\nonum
   &\quad +\frac{(1+\nu ) d_1 \left((1-\gamma ) \left(f_8^2-f_1^2\right)-(1+\nu )   \left(f_8^2-f_2^2\right)\right)}
    {(1-\nu ) (\gamma +\nu ) \left(d_1-(1-\nu ) c_3\right)} \biggr], \label{eq:mass}\\
  J_\psi &= c^3 J^{\rm vac}_\psi+s^3 J^{\rm vac}_\phi , \label{eq:Jpsi}\\
  J_\phi &= c^3 J^{\rm vac}_\phi + s^3 J^{\rm vac}_\psi \label{eq:Jphi},
\end{align}
with
\begin{align}
  J^{\rm vac}_\psi &= -\frac{ \pi \ell^3\tilde{v}^3_0 \left(c_3 f_4 g_3\left(d_1-d_2\right) +c_2 f_5 \left(c_1 c_3 f_4 f_{16}-b d_1 f_2 f_{13}\right)\right)}
  {8 G_5 \nu  (1-\gamma   )^2 (\gamma^2-\nu^2)  }, \label{eq:Jpsi-vac}\\
  J^{\rm vac}_\phi &= -\frac{\pi  \ell^3\tilde{v}^3_0 \left(  2 \nu  c_3 f_3 f_1 f_{10}-(2 \nu  c_3 f_3 g_{9}+ d_1 f_1   g_{1})(a-b) (1-\gamma ) (1-\nu )\right)}{4 G_5 (1-\gamma ) (1-\nu^2 ) (\gamma^2-\nu^2)}.\label{eq:Jphi-vac}
\end{align}
From the asymptotic behavior of the gauge potential~(\ref{eq:gaugesol}), one can determine the electric charge
\begin{align}\label{eq:Qe}
 Q& := \fr{8\pi G_5} \int_S \left( \star F + \fr{\sqrt{3}} F\wedge A\right)\nonum
 & = \fr{8\pi G_5} \int_{S_\infty} \star F\nonum
 & = - \frac{2M \tanh(2\alpha)}{\sqrt{3}},
\end{align}
where $S$ is an $S^3$-surface that encloses both $\partial \Sigma_{\cal H}$ and $\partial \Sigma_{\rm in}$, and $S_\infty$ represents spatial infinity.
Note that, in the second line, we dropped the Chern-Simons term as it decays faster than the first term at the infinity.

\subsection{Rotational axes ($y=-1,x=\pm 1$)}

\medskip
The boundaries  $\partial\Sigma_\psi$ and $ \partial\Sigma_\phi$ are the $\psi$ and $\phi$ rotational axes where the Killing vectors $\partial_\psi$ and $\partial_\phi$ vanish, respectively.
Moreover, on these axes, there exist no conical singularities, due to the periodicities of $\psi$ and $\phi$ specified in Eq.~(\ref{eq:tphipsirange}).

The inner axis $ \partial\Sigma_{\rm in}$ corresponds to the points where the Killing vector 
$v_{\rm in} = \left(v_{\rm in}^t, v_{\rm in}^{\psi}  ,1 \right)$ vanishes, 
where
\begin{align}
v_{\rm in}^t &= \ell \tilde{v}_0 \left( \frac{c^3 \tilde{g}_1(a-b)}{f_1} - \frac{s^3g_{12}(1-ab)}{f_2}\right),\label{eq:vint}\\
 v_{\rm in}^{\psi} &=\frac{\Delta}{f_1f_2}\left(\frac{ad_1+(1-\gamma)(1+\nu)(1-a^2)c_1}{d_1}-1\right)+1,\label{eq:vinpsi}
\end{align}
with 
\begin{align}
\tilde{g}_1 &:=   f_{13} f_{14}+ \frac{c_3 f_3 f_{10}}{2 (a-b) (1-\gamma ) d_1}-\frac{c_1 c_3\left(f_{13} f_{14}-f_{16}   f_{19}\right)}{2 (a-b) (1-\gamma )^2 \nu },\\
g_{12} &:= f_{5}f_{13}+\frac{c_3(f_4 f_{17}-f_{5}f_{13})}{2\nu (1-\gamma)(1-ab)}.
\end{align}
It is  required that there exist  no Dirac-Misner string singularities or conical singularities, and  the horizon must  have the spherical topology of $S^3$, which is explained as follows:

\medskip

\begin{enumerate}
\item The condition for the absence of Dirac-Misner string singularity: 
The inner axis $\partial \Sigma_{\mathrm{in}}$ is free from Dirac-Misner string singularities if the Killing vector $v_{\mathrm{in}}$ has no time component. 
From Eq.~(\ref{eq:vint}), this condition connects the parameter $\alpha$ with the other parameters: 
\begin{align} \label{eq:def-t3}
\tanh^3 \alpha = \frac{(a-b) f_2 \tilde{g}_1}{(1-ab) f_1 g_{12}}, 
\end{align} 
where the magnitude of the right-hand side must be less than or equal to one.

\item The condition for the horizon topology: 
The cross-section of the Killing horizon has the topology of $L(n;1)$ ($n$: integer) if
\begin{align}
 v_{\rm in}^{\psi} &=\frac{\Delta}{f_1f_2}\left(\frac{ad_1+(1-\gamma)(1+\nu)(1-a^2)c_1}{d_1}-1\right)+1 =n. \label{eq:topology-con}
\end{align}
To specifically ensure a topology of $S^3$, we must set $n = \pm1$.

\item The condition for the absence of conical singularities: 
Since on this axis, the Killing vector $\partial_{\phi_+} := \partial_\phi+n\partial_\psi$ vanishes under the  two conditions previously mentioned, 
the inner axis does not exhibit  a conical singularity if
\begin{align}\label{eq:conifree}
\left(\frac{\Delta \phi_+}{2\pi}\right)^2= \frac{d_1^2 f_1^2 f_2^2}{(1-\gamma)^3(1-\nu)^2(1+\nu)(1-a^2)^2 \Delta^2}=1.
\end{align}

\end{enumerate}

\medskip

\subsubsection{Boundary behavior around $\partial \Sigma_\psi$}
One can show $\partial \Sigma_\psi$ represents a regular rotational axis
in the radial coordinate $r$ by $y=-1-C_0 r^2$ with a positive constant $C_0$.
Then, the metric at $r \to 0$ ($y\to -1$) behaves as
\begin{align}
 ds^2 \simeq \gamma^0_{tt}(x)dt^2 + 2 \gamma^0_{t\phi}(x) dt d\phi + \gamma^0_{\phi\phi}(x)d\phi^2 + \alpha_0(x)
 (dr^2+r^2 d\psi^2+G^{-1}(x)dx^2),
\end{align}
where 
\begin{align}
&\gamma_{tt}^{0} =-\frac{D|_{y=-1}H(-1,x)}{H(x,-1)},\quad
\gamma_{t\phi}^{0} = -\frac{H(-1,x)[c^3 \Omega_\phi(x,-1)-s^3 \Omega_\psi(-1,x)]}{D^2|_{y=-1}H(x,-1)},\nonum
&\gamma_{\phi\phi}^{0} =-\frac{D|_{y=-1} F(x,-1)}{H(-1,x)}-\frac{H(-1,x)[c^3\Omega_\phi(x,-1)-s^3 \Omega_\psi(-1,x)]^2}{D^2|_{y=-1} H(x,-1)},\quad
 \alpha_0 = \frac{ C_0 \ell^2 D|_{y=-1} H(x,-1)}{2(1-\gamma)^3(1-\nu)^3\Delta(1-a^2)(1+x)^2}.
\end{align}
Under the assumptions of Eq.~(\ref{eq:necessary-1}), $H(x,y)> 0$ and $D > 0$,  one can show that $\alpha_0>0$ and 
\begin{align}
{\rm det}\,(\gamma^0)
=- \frac{16\ell^2 (1-\gamma)^3(1 - \nu)^4(1-a^2)\Delta (1-x)(1+\nu x)}{(1+x) D |_{y=-1} H(x, -1)}<0,
\end{align}
meaning that $\gamma^{0}_{AB}(A,B=t,\phi)$ is a nonsingular and non-degenerate matrix for $-1<x<1$. 
Therefore, the metric is regular at $y=-1$.

\medskip

\subsubsection{Boundary behavior around $\partial \Sigma_\phi$ and $\partial \Sigma_{\rm in}$}
Similarly, $\partial \Sigma_\phi$ and $\partial \Sigma_{\rm in}$ can be shown to be the $\phi$-rotational axis and inner rotational axes, respectively.
Using the radial coordinate $r$ by  $x = \pm 1 \mp C_\pm r^2$ with positive constants $C_\pm$ for $x=\pm 1$,
we can see that, with the use of Eqs.~(\ref{eq:def-t3}), (\ref{eq:topology-con}) and (\ref{eq:conifree}), the metric at $r\to0$ ($x\to \pm 1$) behaves as
\begin{align}
 ds^2 \simeq \gamma^{\pm}_{tt}(y) dt^2+2 \gamma^{\pm}_{t\psi}(y) dt d\psi_\pm+ \gamma^{\pm}_{\psi\psi}(y) d\psi_\pm^2 + \alpha_\pm(y) ( dr^2+r^2 d\phi_\pm^2-G^{-1}(y) dy^2),
\end{align}
where
\begin{align}\label{eq:localmetricfunc-x1}
&\gamma_{tt}^{\pm} =-\frac{D|_{x=\pm1}H(y,\pm1)}{H(\pm1,y)},\quad
\gamma_{t\psi}^{\pm} = -\frac{H(y,\pm1)[c^3 \Omega_\phi(\pm1,y)-s^3 \Omega_\phi(y,\pm1)]}{D^2|_{x=\pm1}H(\pm1,y)},\nonum
&\gamma_{\psi\psi}^{\pm} =\frac{D|_{x=\pm1} F(y,\pm1)}{H(y,\pm1)}-\frac{H(y,\pm1)[c^3\Omega_\psi(\pm1,y)-s^3 \Omega_\phi(y,\pm1)]^2}{D^2|_{x=\pm} H(\pm1,y)},\nonum
& \alpha_{\pm} = \frac{C_\pm \ell^2 D |_{x=\pm 1} H(\pm 1,y)}{2(1-\gamma)^3(1\pm \nu)(1-\nu)^2\Delta(1-a^2)(1\mp y)^2},
\end{align}
and
\begin{align}
(\psi_{-},\phi_{-}):=(\psi,\phi),\
(\psi_{+},\phi_{+}):= (\psi-n\phi,\phi).\label{eq:defpsipm}
\end{align}
Under the assumptions $H(x,y)>0$ and $D>0$, we can also show that  $\alpha_{\pm}>0$,  and 
\begin{align}
 &{\rm det} \, (\gamma^{+})
 = \frac{16\ell^2 d_1^2 (1+\nu) ( 1+y)(1+\nu y)}{(1-a^2)(1-y)D |_{x=  1} H( 1,y)}<0,\\
&{\rm det} \, (\gamma^{-})
 = \frac{16\ell^2(1-\gamma)^3(1-\nu)^4(1-a^2)( 1-y)(1+\nu y)}{(1+y)D |_{x= - 1} H( -1,y)}<0,
\end{align}
hence $\gamma^{\pm}_{AB}(A,B=t,\psi)$ is a nonsingular and non-degenerate matrix for $-1/\nu<y<-1$. 
Thus, the metric is regular at $x=\pm1$.

Each point in $\partial \Sigma_{\rm in}$ has the topology of $S^1$ generated by $\psi_+$, which shrinks to a point at $y=-1$ but does not at $y=-1/\nu$. Therefore, $\partial \Sigma_{\rm in}$ has the topology of two-dimensional disk, and called disk-shaped bubble.
One can define the two-dimensional area of the bubble $\cD$ by
\begin{align}\label{eq:def-bubble-area}
A_\cD :=  \int_\cD \sqrt{\alpha_+(y) G^{-1}(y) \gamma^+_{\psi\psi}(y)} d\psi_+  dy.
\end{align}

\subsection{The center $(x,y)=(1,-1)$}

At $(x,y)=(1,-1)$ where the $\psi$-rotational axis and inner rotational axis intersect, the conditions $H(x,y)>0$ and $D>0$ lead to further inequalities 
\begin{align}
&H(x=1,y=-1) = -8(1-\gamma)(1-\nu)^2(1+\nu)d_1 f_2^2 >0 \quad \Longleftrightarrow  \quad  d_1 = (\nu +1) c _1^2-(1-\gamma ) (1-\nu )^2<0,\label{eq:necessary-2}\\
& D(x=1,y=-1) = c^2 - s^2 \frac{(1-\gamma)f_1^2}{(1+\nu) f_2^2} >0  \quad \Longleftrightarrow \quad \tanh^2\alpha < \frac{(1+\nu)f_2^2}{(1-\gamma)f_1^2}.
\label{eq:necessary-3}
\end{align}
For later convenience, instead of Eq.~(\ref{eq:necessary-3}), we consider the stronger condition independent of  $\alpha$
\begin{align}
1< \frac{(1+\nu)f_2^2}{(1-\gamma)f_1^2}. \label{eq:necessary-3b}
\end{align}
Assuming these inequalities and introducing the coordinates $(r,\theta)$ by
 \begin{align}
&  x = 1 - {\cal C}_1 (1+\nu) r^2 \cos^2\theta,\quad y=-1-{\cal C}_1 (1-\nu) r^2 \sin^2\theta,\\
& {\cal C}_1 := -\frac{(1-\gamma)^2(1-a^2)\Delta}{\ell^2 d_1 (c^2 (1+\nu)f_2^2-s^2 (1-\gamma)f_1^2)},
 \end{align}
one can show that, with the use of Eqs.~(\ref{eq:def-t3}), (\ref{eq:topology-con}) and (\ref{eq:conifree}), the metric approaches at $r\to 0$  
\begin{align}
ds^2 \simeq -dt'^2+(\cC_1/|\cC_1|))\left[ dr^2 +r^2(d\theta^2+\sin^2 \theta d\psi_+^2  + \cos^2 \theta d\phi_+^2)\right],
\end{align}
where $t' := \sqrt{(1-\gamma)(1+\nu)}f_1 f_2t/(c^2 (1+\nu)f_2^2-s^2(1-\gamma)f_1^2) $ and $(\psi_+,\phi_+) = (\psi-n\phi,\phi)$. Since the inequalities~(\ref{eq:necessary-1}), (\ref{eq:necessary-2}) and (\ref{eq:necessary-3b}) guarantees $\cC_1>0$, the spacetime metric becomes the Minkowski metric. Therefore, the point $(x,y)=(1,-1)$ turns out to be a regular center of the Minkowski spacetime.

\subsection{Horizon}

The boundary 
$\partial \Sigma_{\cal H}=\{(x,y)|-1<x<1,y=-1/\nu\}$ corresponds to
 a Killing horizon with the surface gravity
\begin{align}
\kappa 
  =   \frac{2 \sqrt{2}\sqrt{1-\gamma}(\gamma-\nu)^{3/2} \sqrt{\nu(1+\nu)} (\gamma+\nu)  }{\ell c_3  \tilde{v}_0^3 \left(c^3 \left(a c_1-1+\nu   \right) f_3 g_3+(a-b) s^3   (\gamma -\nu ) f_4 g_{13}\right)  },
\end{align}
and the horizon angular velocities
 $v_{\cal H}=(1,\omega_\psi,\omega_\phi)$:
\begin{align}
&\omega_{\psi} := -\frac{(\gamma -\nu )  f_3
   f_4}{\ell \tilde{v}_0 \left(c^3 \left(a   c_1-1+\nu \right) f_3 g_3+s^3(a-b)    (\gamma -\nu ) f_4 g_{13}\right)  },\\
&    \omega_{\phi} :=\frac{(\gamma -\nu ) d_3 
   g_2}{\ell \tilde{v}_0 c_3 \left(c^3 \left(a   c_1-1+\nu \right) f_3 g_3+s^3(a-b)    (\gamma -\nu ) f_4 g_{13}\right) }, \end{align}
where
   \begin{align}
   g_{13} := f_8 f_{18}-\frac{f_1 f_{10}}{2 (1-\gamma ) (1-\nu ) (b-a)}.
   \end{align}
The spatial cross section of the horizon has topology $S^3$ under the boundary condition~(\ref{eq:topology-con}) with $n=\pm 1$.
The area of the horizon cross section at $y=-1/\nu$ is given in terms of the surface gravity
\begin{align}
 A_{\cal H} = 8\pi^2 \ell^2 \nu \kappa^{-1}.
\end{align}

Let us show that both the metric and gauge potential are regular at $y=-1/\nu$, by introducing the ingoing/outgoing Eddington-Finkelstein coordinates by
\begin{align}
 dx^i = dx'^i \pm v_{\cal H}^i \frac{(1-\nu^2)}{2 \nu \kappa G(y)} dy, 
\end{align}
where $(x^i)=(t,\psi,\phi)$ $(i=0,1,2)$ and the metric near $y=-1/\nu$ behaves as
\begin{align}
ds^2 &\simeq \alpha_H(x) \left(\frac{4 \nu^2 \kappa^2 G(y) }{(1-\nu^2)^2}dt'^2 \pm \frac{4\nu \kappa}{1-\nu^2} dt' dy+\frac{dx^2}{G(x)} \right)\nonum
&+ \gamma^H_{\psi\psi}(x)(d\psi'-\omega_\psi dt')^2+ 2\gamma^H_{\psi\phi}(x)(d\psi'-\omega_\psi dt')(d\phi'-\omega_\phi dt')
+ \gamma^H_{\phi\phi}(x)(d\phi'-\omega_\phi dt')^2,
\end{align}
with
\begin{align}
\begin{split}
&\gamma_{\psi\psi}^{H} =\frac{D|_{y=-1/\nu} F(-1/\nu,x)}{H(-1/\nu,x)}
-\frac{H(-1/\nu,x)[c^3\Omega_\psi(x,-1/\nu)-s^3 \Omega_\phi(-1/\nu,x)]^2}{D^2|_{y=-1/\nu} H(x,-1/\nu)},\\
&\gamma_{\psi\phi}^{H} =-\frac{D|_{y=-1/\nu} J(x,-1/\nu)}{H(-1/\nu,x)}-\frac{H(-1/\nu,x)[c^3\Omega_\psi(x,-1/\nu)-s^3 \Omega_\phi(-1/\nu,x)][c^3\Omega_\phi(x,-1/\nu)-s^3 \Omega_\psi(-1/\nu,x)]}{D^2|_{y=-1/\nu} H(x,-1/\nu)},\\
&\gamma_{\phi\phi}^{H} =-\frac{D|_{y=-1/\nu} F(x,-1/\nu)}{H(-1/\nu,x)}-\frac{H(-1/\nu,x)[c^3\Omega_\phi(x,-1/\nu)-s^3 \Omega_\psi(-1/\nu,x)]^2}{D^2|_{y=-1/\nu} H(x,-1/\nu)},\\
& \alpha_H =\frac{ \ell^2 \nu^2 D|_{y=-1/\nu}  H\left(x,-1/\nu \right)}{4(1-\gamma)^3(1-\nu)^2\left(1-a^2\right) \Delta(1+\nu x)^2}.
\end{split}
\end{align}
Hence, under the assumptions $H(x,y)>0$ and $D>0$, we can show $\alpha_H>0$ and
\begin{align}
{\rm det}\, (\gamma^H)
= \frac{4 \ell^2 (1-\gamma)^3(1-\nu)^4 (1+\nu)^2 (1-a^2)\Delta(1-x^2)}{\kappa^2D(x,y=-1/\nu) H(x,-1/\nu)(1+\nu x)}
\end{align}
and thus $\gamma^{H}_{AB}(A,B=\psi,\phi)$ is a nonsingular and non-degenerate matrix for $-1<x<1$, meaning that the metric is smoothly continued to $-\infty<y<-1/\nu$ across the horizon $y=-1/\nu$. 
Moreover, in the Eddington-Finkelstein coordinate, the gauge potential also remains regular at the horizon $y=-1/\nu$ under the gauge transformation
\begin{align}
 A' =  A \pm d\left( \frac{(1-\nu^2)\Phi_e}{2\nu \kappa} \int \frac{dy}{G(y)}\right),
\end{align}
where $\Phi_e$ is the electric potential defined by
\begin{align}
\Phi_e &:= - i_{v_{\cal H}} A \bigr|_{y=-1/\nu}  
=- (A_t + \omega_\psi A_\psi + \omega_\phi A_\phi)\bigr|_{y=-1/\nu}\nonum
& =-\frac{\sqrt{3} c s \left(f_4 g_{13} s (b-a) (\gamma -\nu )-c f_3 g_3 
\left(a c_1+\nu -1\right)\right)}{f_4 g_{13} s^3 (b-a) (\gamma -\nu )-c^3 f_3 g_3 \left(a c_1+\nu   - 1\right)}.
\label{eq:phi-e}
\end{align}

\medskip

The points $(x,y)=(\pm 1,-1/\nu)$, at which the rotational $\phi$ or  $\phi_+(= \phi)$ axes and the horizon intersect, represent  merely regular points.
By introducing the coordinates $(r,\theta)$ for $(x,y)=(\pm 1,-1/\nu)$ as
\begin{align}
  x = \pm 1 \mp c_\pm r^2 \sin^2\theta,\quad y =-\fr{\nu} \left(1- \frac{(1\mp\nu)c_\pm}{2} r^2 \cos^2\theta\right),\quad c_\pm : {\rm positive \ constants},
\end{align}
and with the use of Eqs.~(\ref{eq:topology-con}) and (\ref{eq:conifree}),
the metric at $r\to 0$ ($(x,y)\to(\pm1,-1/\nu)$) behaves as
\begin{align}
 ds^2 \simeq dr^2 + r^2 d\theta^2 + r^2 \sin^2\theta (d\phi_\pm-\omega_\phi dt)^2  - r^2 \kappa^2 \cos^2 \theta dt^2 + R_{\pm}^2 (d\psi_\pm-\omega_\psi^\pm dt)^2,
\end{align}
where $\omega_\psi^+=\omega_\psi-n\omega_\phi$
, $\omega_\psi^- =\omega_\psi$
, and $(\psi_\pm, \phi_\pm)$ are defined in Eq.~(\ref{eq:defpsipm}).
$R_{\pm}:=\sqrt{g_{\psi\psi}}|_{(x,y)=(\pm1,-1/\nu)}$ are given by
\begin{align}
& R_+ =-\frac{2 d_1 \ell (\gamma -\nu ) \sqrt{\nu  (\gamma +\nu )} \left(b f_1 f_{13} f_{15} s^3 \left(d_2-(1-\gamma ) (1-\nu ) (\gamma +\nu )\right)+c^3 d_2
   f_2 f_{11} f_{14}\right)}{\sqrt{\left(1-a^2\right) (1-\gamma )\Delta} \left(-c^2 d_2^2 f_{14}^2 (\gamma -\nu )-2 f_{15}^2 \nu  s^2
   \left(d_2-(1-\gamma ) (1-\nu ) (\gamma +\nu )\right){}^2\right)},\label{eq:defRplus}\\
&R_- =\frac{2 \ell \sqrt{\nu  (\nu +1) (\gamma +\nu )} \left(f_4 g_{13} s^3 (b-a) (\gamma -\nu )+c^3 f_3 g_3 \left(-a c_1-\nu
   +1\right)\right)}{\sqrt{\left(1-a^2\right) \Delta} \left(2 c^2 (1-\gamma ) f_3^2 \nu +f_4^2 (\nu +1) s^2 (\gamma -\nu )\right)},   \label{eq:defRminus}
\end{align}
and we also set $c_{\pm}$ as 
\begin{align}
 c_\pm = \frac{(1\pm\nu)R_\pm\kappa}{2\ell^2 \nu(1\mp\nu)}.
 \end{align}
In the Cartesian coordinates $(T, X, Y, Z,W)=(\kappa t, r\cos\theta, r\sin\theta \cos(\phi_\pm-\omega_\phi t), r\sin\theta \sin(\phi_\pm-\omega_\phi t),R_{\pm} (\psi_\pm-\omega^\pm_\psi t) )$, the above asymptotic metric becomes
\begin{align}
 ds^2 \simeq - X^2 dT^2 + dX^2 + dY^2 + dZ^2 +dW^2,
 \end{align}
 where the Rindler horizon lies at $X=0$.
Therefore, the metric is regular at $(x,y)=(\pm1,-1/\nu)$.

\subsection{Absence of curvature singularities and CTCs}

In the previous subsections, we have derived the conditions ensuring the regularity of the solution at the boundaries of the C-metric coordinates $(x,y)$. 
Here, we demonstrate that these conditions are also sufficient to guarantee the solution's regularity throughout the entire region spanned by the C-metric coordinates, as specified in Eq.~(\ref{eq:xyrange}). 
We show that under the inequalities~(\ref{eq:necessary-1}), (\ref{eq:necessary-2}) and (\ref{eq:necessary-3b}), the conditions~(\ref{eq:def-t3}), (\ref{eq:topology-con}) and (\ref{eq:conifree}) ensures $H(x,y)>0$ and $D>0$ within the region specified in~(\ref{eq:xyrange}).

\medskip

\begin{figure}
\includegraphics[width=6cm]{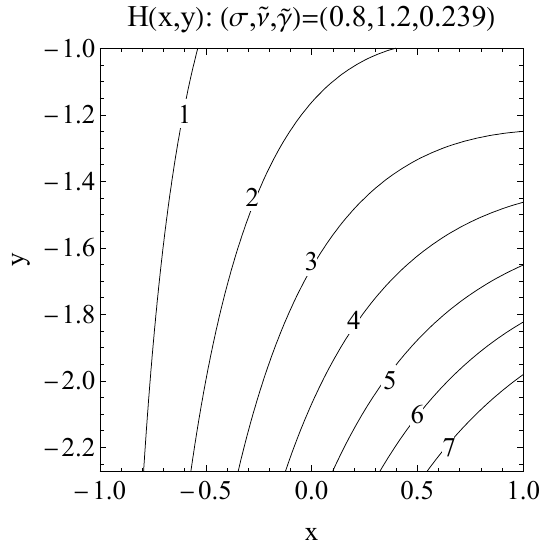}
\caption{Profile of $H(x,y)$ for $(x,y) \in [-1, 1]\times [-1/\nu , -1]$ and $(\sigma,\tilde{\nu},\tilde{\gamma})=(0.8,1.2,0.239)$ with $n=1$.\label{fig:hxy}}
\end{figure}

\medskip
It is challenging to demonstrate analytically $H(x,y)>0$,  however, as depicted in Fig.~\ref{fig:hxy}, we can numerically confirm $H(x,y)>0$ for the parameters satisfying the conditions~(\ref{eq:def-t3}), (\ref{eq:topology-con}) and (\ref{eq:conifree}), as well as the  inequalities (\ref{eq:necessary-1}), (\ref{eq:necessary-2}) and (\ref{eq:necessary-3b}). 
Once $H(x,y)>0$ is established, it follows that $D>0$ within the ranges (\ref{eq:xyrange}) because of the relation $D=1+s^2(H(x,y)-H(y,x))/H(x,y)\geq1$ and the inequality 
\begin{align}
  H(x,y)-H(y,x) &=(x-y)\biggr[\frac{(1-\nu ) \left((\gamma -\nu ) d_2^2 f_{14}^2
  +2 \nu  \left(d_2-(1-\gamma ) (1-\nu ) (\gamma +\nu   )\right){}^2 f_{15}^2\right)}{\gamma -\nu }(1+x)(-1-y)\nonum
& + \frac{4 (1-\gamma ) (1-\nu ) (-d_1)(1+\nu ) f_2^2}{1+\nu } \left(1-\frac{(1-\gamma ) f_1^2}{(1+\nu ) f_2^2}\right)(1+\nu x)(1+\nu y)\nonum
& +\frac{(1-\gamma ) (1-\nu )^3 c_3 \left(2 (1-\gamma ) \nu  f_3^2+(\gamma -\nu ) (1+\nu ) f_4^2\right)}{1+\nu
   } (1-x)(1-y) \biggr] \geq 0,
\end{align}
where nonnegativity of the second term follows from Eq.~(\ref{eq:necessary-3b}).

\medskip
The absence of closed timelike curves (CTCs) is demonstrated if the two-dimensional metric $g_{IJ} (I,J=\psi,\phi)$ is positive definite, except at the axes where $x=\pm1$ and $y=-1$. This corresponds to the conditions $\mathrm{det}(g_{IJ})>0$ and $\mathrm{tr}(g_{IJ})>0$ for $-1<x<1$ and $-1/\nu \leq y <-1$. Following the reasoning discussed in Ref.~\cite{Suzuki:2024coe}, demonstrating that $\mathrm{det}(g_{IJ})>0$ is sufficient.

\medskip

$\mathrm{det}(g_{IJ})$ vanishes on the $\phi$ and $\psi$-rotational axes at $x=-1$ and $y=-1$, as well as on the inner axis at $x=1$ after imposing Eq.~(\ref{eq:def-t3}).
To demonstrate the positive definiteness of $\mathrm{det}(g_{IJ})$ away from these, 
 we consider the positive definiteness of the alternative function ${\cal D} (x,y)$ for $-1 \leq x \leq 1$ and $-1/\nu \leq y \leq -1$, defined by
\begin{align}
{\cal D} (x,y):=\frac{(x-y)^4 D H(x,y)}{\ell^4 \tilde{v}_0^4(1-x^2)(-1-y)}  {\rm det}(g_{IJ}).
\end{align}
Although giving the analytic proof for the positivity of ${\cal D} (x,y)$ in the entire region is challenging, one can still possible to demonstrate positivity on the horizon:
\begin{align}
{\cal D} (x,-1/\nu) = \frac{(1-\gamma ) c_3^2 (1-\nu )^3 (\nu +1) (1+\nu  x)^3 \left[c^3 f_3 g_3 \left(-a c_1-\nu +1\right)-f_4 g_{13} s^3 (a-b) (\gamma -\nu )\right]{}^2}{\nu ^4
   (\gamma -\nu )^2 (\gamma +\nu )} >0.
\end{align}
\begin{figure}
\includegraphics[width=6cm]{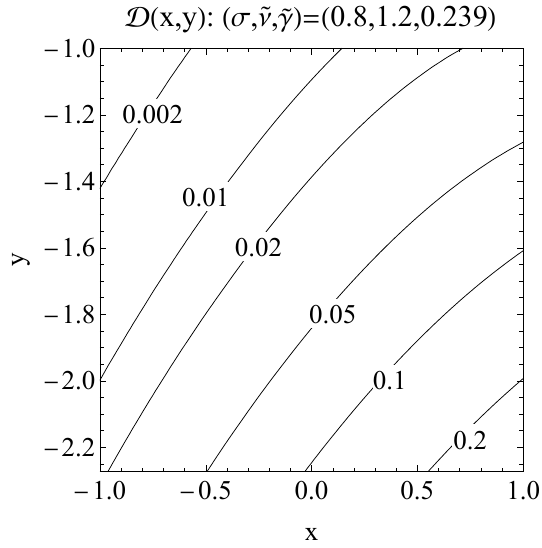}
\caption{Profile of ${\cal D}(x,y)$ for $(x,y)\in [-1, 1] \times [-1/\nu , -1]$ and $(\sigma,\tilde{\nu},\tilde{\gamma})=(0.8,1.2,0.239)$ with $n=1$.\label{fig:dxy}}
\end{figure}
For the other region,  as shown in Fig.~\ref{fig:dxy}, 
we numerically confirmed the positivity both on and outside the horizon for several choices of the parameters under the constraints~(\ref{eq:def-t3}), (\ref{eq:topology-con}) and (\ref{eq:conifree}) followed by the inequalities~(\ref{eq:necessary-1}), (\ref{eq:necessary-2}) and (\ref{eq:necessary-3}).

\medskip

\subsection{Finding parameters for capped black holes}

\medskip

We are now aiming to find a capped black hole solution that does not have any curvature singularities, orbifold singularities, conical singularities, or Dirac-Misner string singularities  both on and outside the horizon, additionally, they are free from closed timelike curves. 
This can be achieved  if  we can find a set of parameters $(\ell,\nu,\gamma,a,b,\beta,\alpha)$ that meets three constraints~(\ref{eq:def-t3}), (\ref{eq:topology-con}) and (\ref{eq:conifree}).
As outlined in our previous works~\cite{Suzuki:2023nqf, Suzuki:2024phv}, such black holes are distinguished by having a compact inner rotational axis with a two-dimensional disk topology, referred to as a ``bubble," which lies adjacent to the horizon. 
The existence of this extra bubble implies that the intersection of the DOC and timeslice $\Sigma$ has the nontrivial topology of ${\rm DOC}\cap \Sigma = [{\mathbb R}^4 \# {\mathbb C} {\mathbb P}^2 ]\setminus {\mathbb B}^4$, differing it from the Cveti\v{c}-Youm black hole with the trivial topology of ${\rm DOC}\cap \Sigma = {\mathbb R}^4 \setminus {\mathbb B}^4$~\cite{Cvetic:1996xz}.
We analyze the cases of $n=1$ and $n=-1$ separately.

\medskip
\paragraph{Capped black hole with $n=1$.}

To solve the constraints, we introduce the new parameters as
\begin{align}
\tilde{\nu} := \sqrt{1+\nu},\quad \tilde{\gamma}:=\sqrt{1-\gamma},
\label{eq:def-paramtilde}
\end{align}
which is helpful in avoiding the square root terms in the later expressions.
From Eq.~(\ref{eq:rangenugam}), they are restricted to the range
\begin{align}\label{eq:new}
1<\tilde{\nu}<\sqrt{2},\quad 0<\tilde{\gamma}<\sqrt{2-\tilde{\nu}^2}.
\end{align}
Then, Eq.~(\ref{eq:topology-con}) can be solved with respect to $a$ as
\begin{align}
a = \frac{\tilde{\gamma }^2 \left(\tilde{\nu }^2-2\right) \left(\left(2 b^2+b-1\right) \tilde{\nu }^2+2\right)+b^2 \tilde{\nu }^2 \left(\tilde{\nu }^2-2\right)^2+b (b+1)
   \tilde{\gamma }^4 \tilde{\nu }^2}{\tilde{\gamma }^2 \tilde{\nu }^2 \left((b+1) \tilde{\gamma }^2+b \left(\tilde{\nu }^2-2\right)\right)}.\label{eq:sola}
\end{align}
Using Eqs.~(\ref{eq:def-paramtilde}) and (\ref{eq:sola}), one can rewrite Eq.~(\ref{eq:conifree}) as
\begin{align}
&\left[\Delta \left(\left(b^2-1\right) \tilde{\gamma }^4 \tilde{\nu }^2+\tilde{\gamma }^2 \left(\tilde{\nu }^2-2\right) \left(\left(2 b^2-1\right) \tilde{\nu }^2+2\right)+b^2
   \tilde{\nu }^2 \left(\tilde{\nu }^2-2\right)^2\right)+f_1 f_2 \tilde{\gamma }^3 \tilde{\nu } \left(\tilde{\nu }^2-2\right)\right]\nonum
&\times \left[ \Delta \left(\left(b^2-1\right) \tilde{\gamma }^4 \tilde{\nu }^2+\tilde{\gamma }^2 \left(\tilde{\nu }^2-2\right) \left(\left(2 b^2-1\right) \tilde{\nu }^2+2\right)+b^2
   \tilde{\nu }^2 \left(\tilde{\nu }^2-2\right)^2\right)-f_1 f_2 \tilde{\gamma }^3 \tilde{\nu } \left(\tilde{\nu }^2-2\right)\right]=0.  \label{eq:norm-n1p}
\end{align}
Since, with the use of  Eq.~(\ref{eq:sola}), $f_1,f_2$ and $\Delta$ is invariant under the transformation $\tilde{\gamma} \to - \tilde{\gamma}$, Eq.~(\ref{eq:norm-n1p})
reduces to a single condition by letting $\tilde{\gamma}$ to take negative values
\begin{align}
\Delta \left(\left(b^2-1\right) \tilde{\gamma }^4 \tilde{\nu }^2+\tilde{\gamma }^2 \left(\tilde{\nu }^2-2\right) \left(\left(2 b^2-1\right) \tilde{\nu }^2+2\right)+b^2
   \tilde{\nu }^2 \left(\tilde{\nu }^2-2\right)^2\right)+f_1 f_2 \tilde{\gamma }^3 \tilde{\nu } \left(\tilde{\nu }^2-2\right)=0. \label{eq:norm-n1p2}
\end{align}
Moreover, with Eq.~(\ref{eq:sola}), the condition~(\ref{eq:necessary-2}) is simplified as
\begin{align}
(d_1 =) -\tilde{\gamma}^2(2-\tilde{\nu}^2)^2(1-\tilde{b}^2)<0 \quad \Longleftrightarrow \quad -1<\tilde{b}<1,
\end{align}
where another new parameter $\tilde{b}$ is introduced by
\begin{align}
 b =-\frac{\tilde{\gamma } \left(\tilde{b} \tilde{\gamma } \tilde{\nu }-\tilde{\nu }^2+2\right)}{\tilde{b} \tilde{\nu } \left(\tilde{\gamma }^2+\tilde{\nu }^2-2\right)}.
\label{eq:def-paramtilde-2}
\end{align}
By substituting Eq.~(\ref{eq:def-paramtilde-2}),
 Eqs.~(\ref{eq:sola}) and (\ref{eq:norm-n1p2}) can be written in terms of the new parameters $(\tilde{\nu},\tilde{\gamma},\tilde{b})$ as
\begin{align}
\Delta \left((\tilde{b}^2-1) \left(2-\tilde{\nu }^2\right)-2 \tilde{b} \tilde{\gamma } \tilde{\nu }\right)+f_1 f_2 \tilde{b}^2 \tilde{\gamma }
   \tilde{\nu }=0,  \label{eq:norm-n1}
\end{align}
\begin{align}
a = \frac{(1-\tilde{b}^2) \left(\tilde{\nu }^2-2\right)-\tilde{b} \tilde{\gamma } \tilde{\nu }}{\tilde{b} \tilde{\gamma } \tilde{\nu }}.
\end{align}
Additionally, to simplify the expression, we replace the parameter $\alpha$ with 
\begin{align}\label{eq:sigma-to-tanh}
\sigma:=\tanh \alpha,\quad -1 < \sigma <1,
\end{align}
where Eq.~(\ref{eq:def-t3}) is written as
\begin{align}\label{eq:def-sigma}
\sigma^3 = \frac{\tilde{b} \tilde{\nu } (\tilde{b} \tilde{\gamma } \tilde{\nu }+\tilde{b}^2 \left(\tilde{\gamma }^2-2\right)+(\tilde{b}^2-1) \tilde{\nu
   }^2+2)}{\tilde{\gamma } (\tilde{b} (\tilde{b}^2-2) \tilde{\gamma } \tilde{\nu }+(1-2 \tilde{b}^2) \tilde{\nu }^2+2
   (\tilde{b}^2-1))} \frac{f_2 \tilde{g}_1}{f_1 g_{12}}.
\end{align}
Note that, from Eq.~(\ref{eq:Qe}), $\sigma$ is directly related to the charge-to-mass ratio as
\begin{align}\label{eq:qtom-sigma}
\frac{Q}{M} = - \frac{4\sigma}{\sqrt{3}(1+\sigma^2)},
\end{align}
where, in particular, the limit $\sigma\to \pm 1(\alpha \to \pm \infty)$ corresponds to the BPS limit $|Q|/M \to 2/\sqrt{3}$.

\medskip
\paragraph{Capped black hole with $n=-1$.}

For $n=-1$, one can eliminate $\Delta,f_1$ and $f_2$ in Eq.~(\ref{eq:conifree}) using Eq.~(\ref{eq:topology-con}) with $n=-1$, which leads to the condition
\begin{align}
\frac{(d_1-(1+a)(1-\gamma)(1+\nu) c_1)^2 }{4(1-\gamma)^3(1-\nu)^2(1+\nu)  (1+a)^2}=1.
\end{align}
In terms of $\tilde{\nu}$ and $\tilde{\gamma}$ defined in Eq.~(\ref{eq:def-paramtilde}), this is solved with respect to $a$ as
\begin{align}
a = a_\pm : = \frac{\tilde{\gamma }^2 \left(\tilde{\nu }^2-2\right) \left(\left(2 b^2+b-1\right) \tilde{\nu }^2+2\right)+b^2 \tilde{\nu }^2 \left(\tilde{\nu }^2-2\right)^2+b (b+1) \tilde{\gamma }^4 \tilde{\nu   }^2\mp 2 \tilde{\gamma }^3 \tilde{\nu } \left(\tilde{\nu }^2-2\right)}{\tilde{\gamma }^2 \tilde{\nu } \left((b+1) \tilde{\gamma }^2 \tilde{\nu }+b \tilde{\nu } \left(\tilde{\nu }^2-2\right)\pm 2   \tilde{\gamma } \left(\tilde{\nu }^2-2\right)\right)}.\label{eq:def-apm-nm1}
\end{align}
Because $a_-$ is derived from $a_+$ by the transformation $\tilde{\gamma}\to -\tilde{\gamma}$, it is sufficient to choose $a=a_+$ without loss of generality, by allowing $\tilde{\gamma}$ to take negative values as in the $n=1$ case.
Substituting $a=a_+$ into the condition~(\ref{eq:necessary-2}), one has 
\begin{align}
d_1 = -\frac{3 (1-\tilde{b}^2 ) \tilde{\gamma }^2 \left(\tilde{\nu }^2-2\right)^2}{(\tilde{b}+2)^2}<0 
\quad \Longleftrightarrow \quad -1<\tilde{b}<1,
\end{align}
where $\tilde{b}$ is defined by
\begin{align}
b =-\frac{\tilde{\gamma } \left(\tilde{\gamma } \tilde{\nu }+\tilde{b} (2-\tilde{\nu }^2) \right)}{\tilde{\nu } 
\left(\tilde{\gamma }^2+\tilde{\nu }^2-2\right)}.\label{eq:def-btildenm1}
\end{align}
Note that $\tilde{b}$ is defined differently from Eq.~(\ref{eq:def-paramtilde-2}) in the $n=1$ case to ensure $\tilde{b}$ maintains the same parameter range
With Eq.~(\ref{eq:def-btildenm1}), $a=a_+$ in Eq.~(\ref{eq:def-apm-nm1}) and Eq.~(\ref{eq:def-t3}) reduce to
\begin{align}
&a = \frac{(\tilde{b}^2-1) (\tilde{\nu }^2-2 )}{(\tilde{b}+2) \tilde{\gamma } \tilde{\nu }}-1,\label{eq:def-newa-nm1}\\
&\sigma^3 = -\frac{ \tilde{\nu } \left(\left(\tilde{b}+2\right) \tilde{\gamma } \tilde{\nu }+\left(2 \tilde{b}+1\right) \tilde{\gamma }^2-\left(\tilde{b}^2-1\right) \left(\tilde{\nu
   }^2-2\right)\right)}{ \tilde{\gamma } \left(\left(2 \tilde{b}^2+2 \tilde{b}-1\right) \tilde{\gamma } \tilde{\nu }+\left(-\tilde{b}^3+2 \tilde{b}+2\right) \tilde{\nu }^2+2 \tilde{b}
   \left(\tilde{b}^2-1\right)\right)}\frac{\tilde{g}_1 f_2}{f_1 g_{12}},\label{eq:def-newsigma-nm1}
\end{align}
where we replaced $\alpha$ with $\sigma$ defined in Eq.~(\ref{eq:sigma-to-tanh}).
Finally, with Eqs.~(\ref{eq:def-btildenm1}) and (\ref{eq:def-newa-nm1}), Eq.~(\ref{eq:topology-con}) with $n=-1$ becomes
\begin{align}
\Delta \left[2  (\tilde{b}+2) \tilde{\gamma } \tilde{\nu }+ (\tilde{b}^2-1) \left(2-\tilde{\nu }^2\right))\right]-3 f_1 f_2 \tilde{\gamma } \tilde{\nu } =0.\label{eq:norm-nm1}
\end{align}

\begin{figure}[t]
\includegraphics[width=7cm]{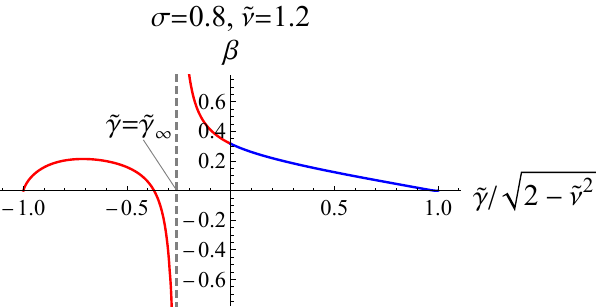}
\includegraphics[width=7cm]{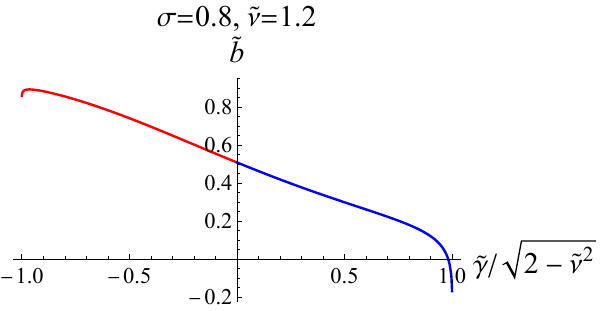}
\caption{All solutions for $n=1$, $\sigma=0.8$ and $\tilde{\nu}=1.2$ shown in the $(\tilde{\gamma},\beta)$ and $(\tilde{\gamma},\tilde{b})$ plains.
$\beta$ diverges at $\tilde{\gamma}_\infty=-0.195...$ and the spacetime seems to be singular there. But the phases are smoothly continued at $\tilde{\gamma}=\tilde{\gamma}_\infty$ if we use the other parameter $\beta'=\beta^{-1}$ around there. 
Despite $\beta=\beta(\tilde{\gamma})$ and $\tilde{b}=\tilde{b}(\tilde{\gamma})$ do not exhibit the mirror symmetry with respect to $\tilde{\gamma}$,
 the metric and gauge functions coincide for $\tilde{\gamma}=C>0$ (blue curve) and $\tilde{\gamma}=-C$ (red curve).  \label{eq:s05phase-params}}
\end{figure}

\medskip
Therefore, to obtain a physical solution, one must solve Eqs.~(\ref{eq:norm-n1}) and (\ref{eq:def-sigma}) for $n=1$ or Eqs~(\ref{eq:def-newsigma-nm1}) and (\ref{eq:norm-nm1}) for $n=-1$ under the conditions~(\ref{eq:necessary-1}) and (\ref{eq:necessary-3}) for the parameters $(\tilde{\gamma},\tilde{\nu},\tilde{b},\sigma,\beta)$ within the specified range
\begin{align}\label{eq:redrange}
 1<\tilde{\nu}<\sqrt{2},\quad 0<|\tilde{\gamma}|<\sqrt{2-\tilde{\nu}^2},\quad -1<\tilde{b}<1,\quad -1<\sigma<1.
\end{align}
For $n=1$, since Eqs.~(\ref{eq:norm-n1}) and (\ref{eq:def-sigma})  yield quartic and sixth-order equations of $\beta$, respectively, 
finding the analytic solution with respect to $\beta$ is not easy. 
Therefore, we resort to exploring the solutions numerically.
The same is true for $n=-1$ with Eqs.~(\ref{eq:def-newsigma-nm1}) and (\ref{eq:norm-nm1}).
Since $\beta$ and $\tilde{b}$ are represented as the functions of $\sigma,\tilde{\nu}$ and $\tilde{\gamma}$ through Eqs.~(\ref{eq:norm-n1}) and (\ref{eq:def-sigma}) for $n=1$ and through Eqs.~(\ref{eq:def-newsigma-nm1}) and (\ref{eq:norm-nm1}) for $n=-1$, 
the physical solution is uniquely determined by a given $(\sigma,\tilde{\nu},\tilde{\gamma})$ in the range~(\ref{eq:redrange}). 
In Fig.~\ref{eq:s05phase-params}, we illustrate all possible solution for $n=1$, $\sigma=0.8,\ \tilde{\nu}=1.2$ and $0<|\tilde{\gamma}|<\sqrt{2-\tilde{\nu}^2}$.
Moreover, we find that the metric and gauge potential is invariant under the transformation $\tilde{\gamma} \to -\tilde{\gamma}$,  
despite $\beta,\tilde{b}$ themselves not being invariant\footnote{The metric function $H(x,y),F(x,y),J(x,y)$ are modified by the transformation $\tilde{\gamma}\to-\tilde{\gamma}$ but  only upto a common factor $H(x,y)\to C H(x,y),\ F(x,y) \to C F(x,y),\ J(x,y) \to CJ(x,y)$. }. As in the black ring case~\cite{Suzuki:2024vzq},
this might imply the possible extension  from the $\tilde{\gamma}=0$ phase into the parameter region $\gamma\  (=1-\tilde{\gamma}^2)>1$ where $\tilde{\gamma}$ becomes pure imaginary and $\beta,\tilde{b}$ become complex. In this paper, however, we restrict ourselves for the phases with all the parameters real.

\medskip

It is worth mentioning  the relationship between the two solutions for $n=1$ and $n=-1$.
Since in a vacuum case, the reflection $x^i \to -x^i$ for either of the Killing coordinates $x^i=(t,\psi,\phi)$ transforms a solution into another with the angular momenta signs flipped,  the transformation do not alter the physical or geometrical properties of the solutions.
However, this is not the case in the five-dimensional minimal supergravity since the Maxwell-Chern-Simons equation~(\ref{Maxeq}) is not invariant under the reflection due to  the presence of the Chern-Simons term.
For example, let us consider the reflection $t\to -t$ for the $t$-component of Eq.~(\ref{Maxeq}): 
\begin{align}\label{Maxeq-2}
\begin{split}
\partial_x (\sqrt{-g} F^{x t} )+\partial_y (\sqrt{-g} F^{yt} ) + \fr{4\sqrt{3}} (F_{\psi x}F_{\phi y}- F_{\phi x}F_{\psi y})=0.
\end{split}
\end{align}
This transformation flips the sign of first two terms, while keeping the last term unchanged, as it results in $F^{xt} \to -F^{xt}, F^{yt} \to -F^{yt}, F_{\psi x}\to F_{\psi x}, F_{\phi x} \to F_{\phi x}$. Similarly,  the other components of Eq.~(\ref{Maxeq}) are not invariant
under $t\to -t$  either.
Instead, one can observe that the following combination of the two reflections 
\begin{align}
 (t,\psi) \to (- t, - \psi) \label{eq:refsym}
\end{align}
 maintains Eq.~(\ref{Maxeq}) invariant. 
Since $g_{\psi\phi}$ and $A_t$ flip the signs, this converts the solution for $n=\pm 1$ and $\sigma$ into the solution for $n_{\rm new}=\mp 1$ and $\sigma_{\rm new}=-\sigma$, where the electric charge also flips,  $Q\to Q^{\rm new}=-Q$.
The mass and angular momenta transform as follows: 
\begin{eqnarray}
(M,J_\psi,J_\phi)\to (M^{\rm new},J_\psi^{\rm new},J_\phi^{\rm new})= (M,J_\psi,-J_\phi)
\end{eqnarray}
because $g_{t\phi}$ flips the sign while $g_{t\psi}$ remains unchanged. 
Therefore, the $n=1$ and $n=-1$ phases  physically represent the same phase through the coordinate transformation~(\ref{eq:refsym}).
As previously mentioned, while all parameters remain real, both $ n=1$ and $n=-1$ phases are incomplete and lack the region for $\gamma>1$.
Nevertheless, it turns out that the transformation~(\ref{eq:refsym}) maps the phase for $n=\pm1$ with real parameters to the phase for $n=\mp1$ in the region including the missing $\gamma>1$ phase (Fig~\ref{fig:extension}).
Hence, in the following section, we will focus on the $n=1$ phase but also use the solution for $n=-1$ to compensate the phase space for $\gamma>1$.

\section{Physical properties}\label{sec:phase}

Finally, we study the phase space of the new solution.
The new capped black hole solution is determined by four independent parameters $(\ell,\sigma,\tilde{\nu},\tilde{\gamma})$. 
Here, we focus on the phase diagram for the $n=1$ case with $\sigma=0.8$. 
Other cases with $\sigma\not=0.8$ are qualitatively similar.

\subsection{Thermodynamics}

From Ref.~\cite{Kunduri:2013vka}, 
the capped black hole follows the first law, which involves local quantities on the bubble $\cD$
\begin{align}
 \delta M = \frac{\kappa }{8\pi} \delta A_{\cal H}  + V_\psi \delta J_\psi + V_\phi \delta J_\phi
 + \fr{2} \Phi_{\cal H} \delta Q + {\cal Q}_\cD \delta \Phi_\cD,
\end{align}
and the Smarr formula
\begin{align}
 M = \frac{3\kappa A_{\cal H}}{16\pi}  + \frac{3}{2} V_\psi  J_\psi + \frac{3}{2}V_\phi J_\phi
 + \fr{2} \Phi_{\cal H} Q  + \fr{2} {\cal Q}_\cD\Phi_\cD,
\end{align}
where $\Phi_{\cal H}$ is the electric potential~(\ref{eq:phi-e}), $\Phi_\cD$ and ${\cal Q}_\cD$ are the magnetic potential and magnetic flux  on $\cD$ defined as
\begin{align}
 \Phi_\cD& := - i_{v_{\rm in}} A \biggr|_{x=1} = -(A_\phi+n A_\psi)\biggr|_{x=1}  =-\frac{\sqrt{3} \ell c s \tilde{v}_0 \left(2 c (1-\gamma ) f_2 \nu  \tilde{g}_1 (a-b)-f_1 s \left(c_3 f_4 f_{17}-b c_2 f_5 f_{13}\right)\right)}{2 \nu (1-\gamma ) 
   f_1 f_2 }, \\
 {\cal Q}_\cD &:= \fr{4}\int_\cD \left[i_{v_H} \star F- \fr{\sqrt{3}} (\Phi-\Phi_{\cal H}) F\right]  =\sqrt{\frac{3}{2}} \frac{ \pi  \ell s  c b d_1 f_{11} f_{13}  \sqrt{\left(1-a^2\right) \Delta } \sqrt{\gamma ^2-\nu ^2}}{2 \sqrt{1-\gamma } c_3 \left(c^3 f_3 g_3
   \left(a c_1+\nu -1\right)-f_4 g_{13} s^3 (b-a) (\gamma -\nu )\right)},
\end{align}
with $\Phi(y) :=- i_{v_{\cal H}} A |_{x=1} = - (A_t + \omega_\psi A_\psi + \omega_\phi A_\phi) |_{x=1}$.

\begin{figure}[t]
\includegraphics[width=13cm]{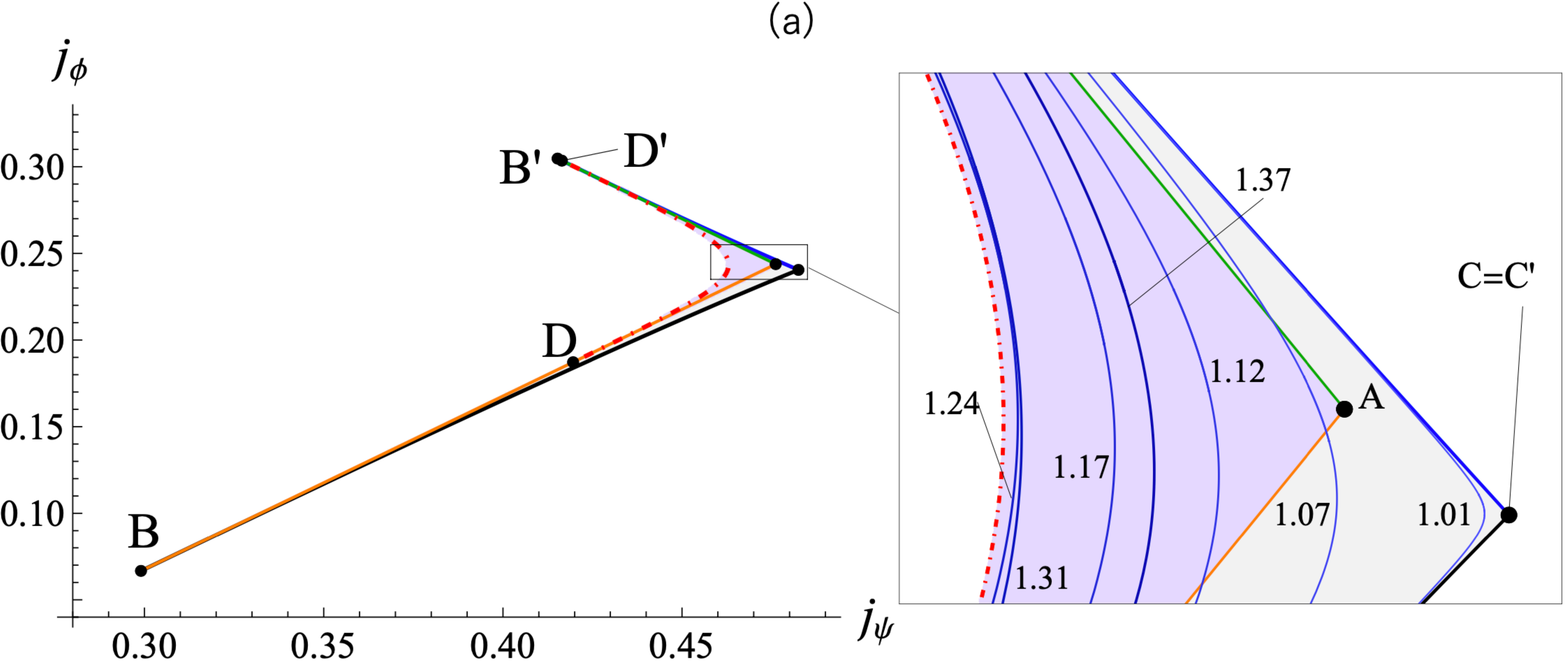}
\vspace{4mm}\\
\includegraphics[width=7cm]{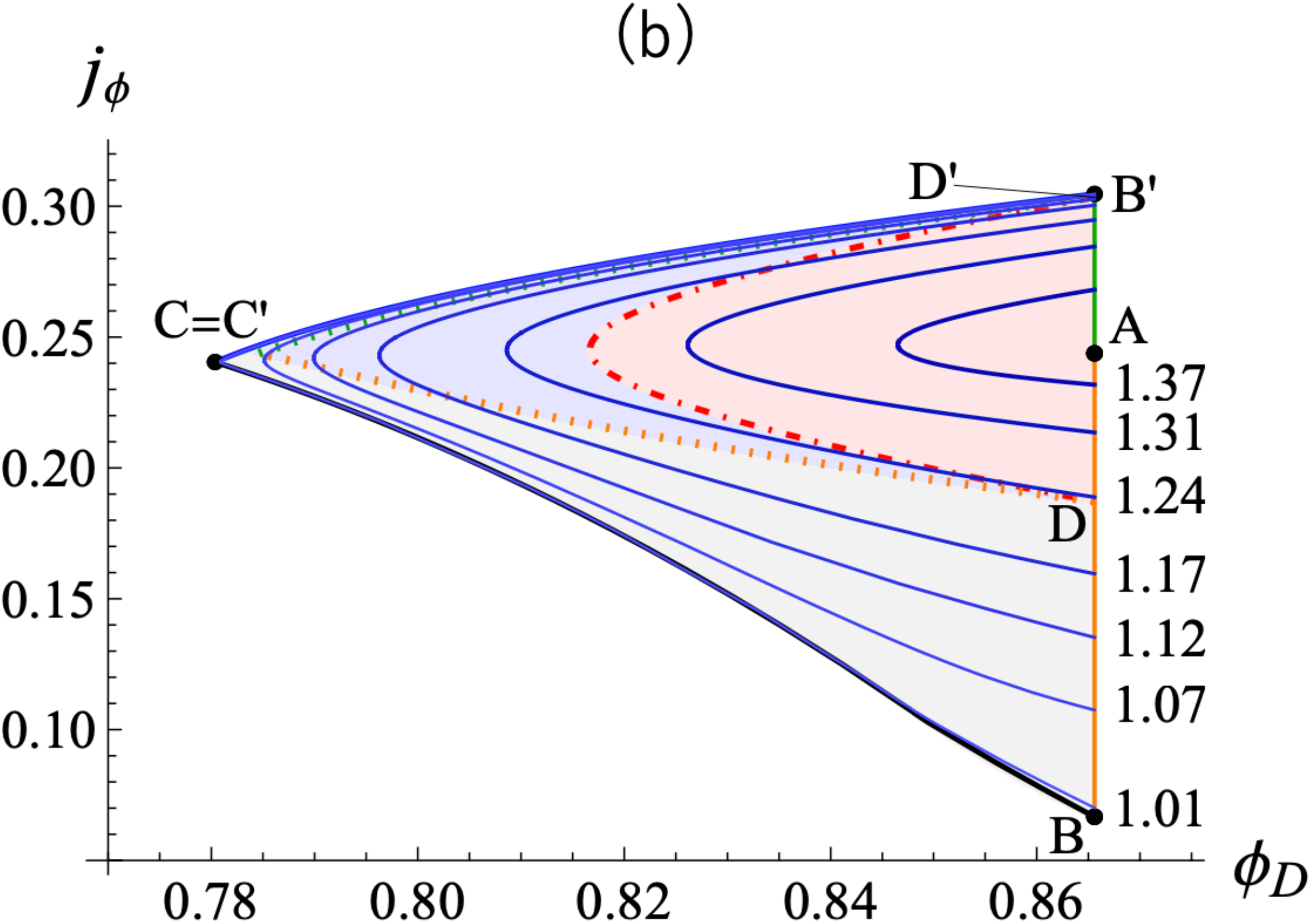}\hspace{2mm}
\includegraphics[width=7.2cm]{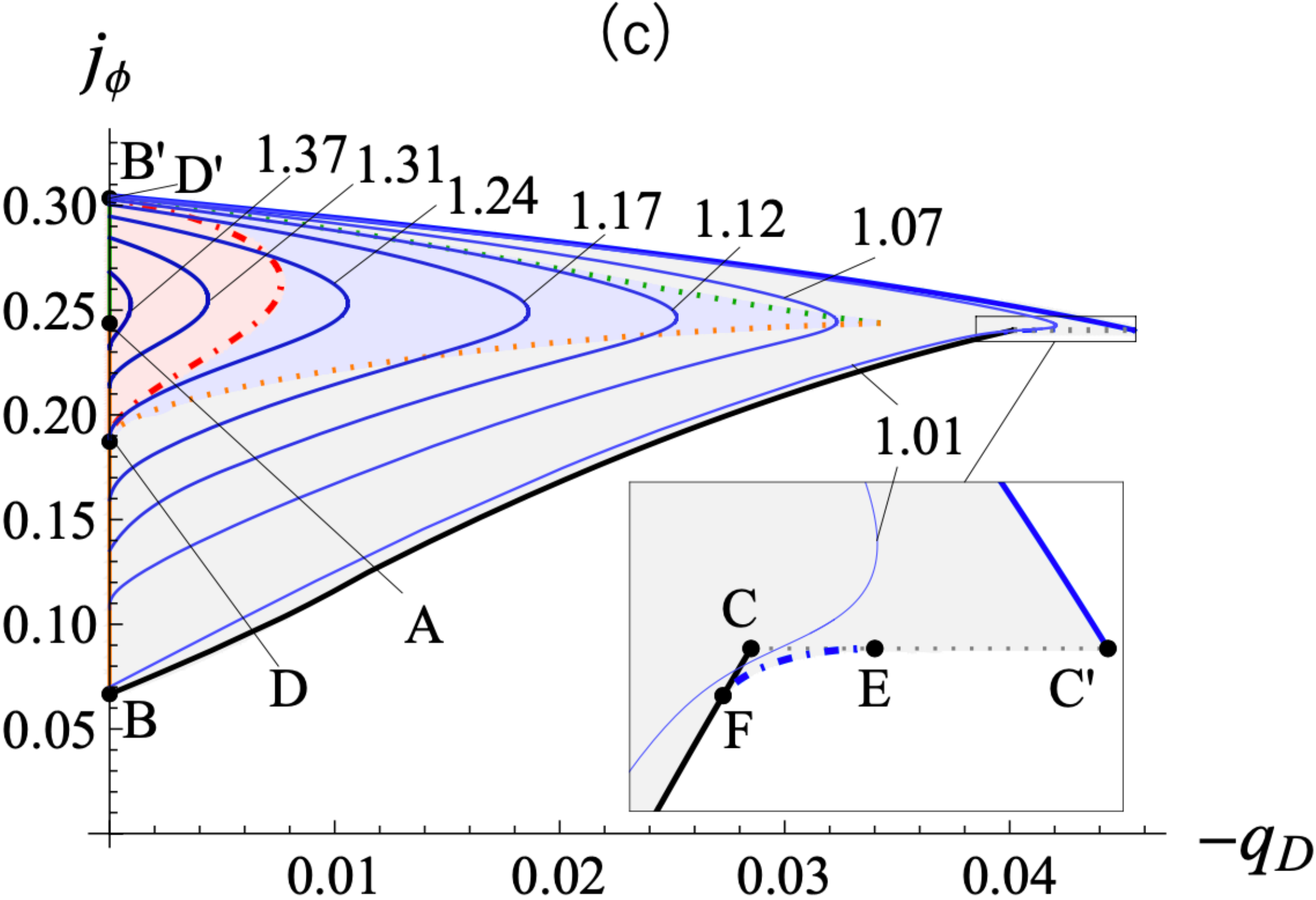}
\caption{
The $\sigma=0.8$ phase in the $(j_\psi,j_\phi)$, $(\phi_\cD,j_\phi)$ and $(q_\cD,j_\phi)$ planes. 
The $\tilde{\nu}={\rm constant}$ phases are shown by thin blue curves. 
To keep the visibility in the panel (a), the $\tilde{\nu}={\rm constant}$ curves are displayed only in the close-up.
In each panel, the limits to the extremal capped black hole are depicted with black and blue curves. 
The limits to the extremal black hole without a bubble are represented by orange and green curves. 
Additionally, the critical curves for the large and small bubble branches are indicated by red dot-dashed lines, which also form part of the boundary in panel (a).
 In panel (a), two branches exist for given $(j_\psi,j_\phi)$ within the purple region, while only the large bubble branch is present in the gray region. 
 In panels (b) and (c), the critical curves separate the two branches, with the orange and green dotted curves marking the transition between regions with single and double branches. 
 The red regions denote the small bubble branch and the blue regions correspond to the overlapping large bubble branch from the purple region in panel (a). 
 In both panels (b) and (c), only the large bubble branch exists within the gray areas.
In each panel,  the two limit curves to the extremal capped black hole and extremal bubbleless black hole  intersect at  points $B$ and $B'$. 
The critical curve terminates at points $D$ and $D'$ on the extremal bubbleless black hole limits. 
In panels (a) and (b), the two limits to  extremal capped black hole join at the point $C=C'$, while,
in panel (c), two limits terminate at different points $C$ and $C'$ linked by the $j_\phi=0.240\ldots$ line (gray dashed line).
Additionally, the phase diagram in panel (c) is delineated by the $\partial_{\tilde{\nu}} j_\phi \partial_{\tilde{\gamma}} q_\cD - \partial_{\tilde{\nu}} q_\cD \partial_{\tilde{\gamma}} j_\phi=0$ curve (blue dot-dashed curve), which begins at $F$ on the lower extremal capped black hole limit and ends at $E$ on the $j_\phi=0.240\ldots$ line.
 \label{fig:phases05} }
\end{figure}

\subsection{Phase diagram for $\sigma=0.8$}

To study the phase diagram of the capped black holes, we introduce the dimensionless variables as follows: 
\begin{align}
j_\psi = \frac{4G_5 J_\psi}{\pi r_M^3},\quad
j_\phi = \frac{4G_5 J_\phi}{\pi r_M^3},\quad
a_{\cal H} = \frac{\sqrt{2}A_{\cal H}}{\pi^2 r_M^3},\quad
\phi_\cD := \frac{\Phi_\cD}{r_M},\quad  q_\cD := \frac{{\cal Q}_\cD}{r_M}.
\end{align}
where $r_M:=8G_5M/(3\pi)$.

\begin{figure}
\includegraphics[width=7.5cm]{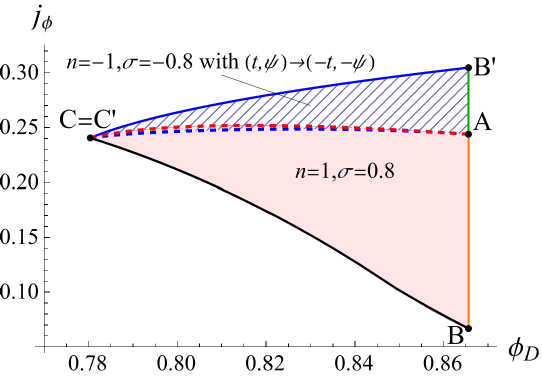}
\caption{
The extension of the $n=1,\sigma=0.8$ phase by using the $n=-1,\sigma=-0.8$ phase.
 The red region represents the $n=1,\sigma=0.8$ phase within the parameter bounds $1\leq \tilde{\nu} \leq \sqrt{2}$ and $0\leq \tilde{\gamma}\leq \sqrt{2-\tilde{\nu}^2}$. 
 The blue hatched region illustrates  the phases derived from the $n=-1,\sigma=-0.8$ phase, for the same ranges of $\tilde{\nu}$ and $\tilde{\gamma}$,  obtained by the transformation $(t,\psi)\to(-t,-\psi)$. 
 The upper boundary of the former region (indicated by a red dashed curve) and the lower boundary of the latter (represented by a blue dashed curve) are defined by the limit $\tilde\gamma\to0$ in each respective solution. Notably, there is a slight overlap between these two regions within these boundaries.
\label{fig:extension}}
\end{figure}

\medskip
In Fig.~\ref{fig:phases05}, the phase for $n=1$ and $\sigma=0.8$ is depicted in the $(j_\psi,j_\phi)$, $(\phi_\cD,j_\phi)$, and $(q_\cD,j_\phi)$ planes, where the solution exists within the colored regions. 
Since the BPS bound is saturated only at the limit $\sigma \to \pm 1$, all the phases for $\sigma=0.8$ are non-BPS (non-supersymmetric).
In each panel, curves of constant $\tilde{\nu}$ are shown as thin blue lines. 
The capped black hole solutions are not uniquely determined by their conserved charges $(M,Q,J_\psi,J_\phi)$ and feature two distinct branches, which we call {\it large bubble branch} and {\it small bubble branch}. 
As explained in detail below,  the former always has the larger area of a bubble than the latter. 
In Fig.~\ref{fig:phases05}(a), both branches coexist within the purple region. 
These branches are distinguishable by their magnetic flux or potential values. 
In Figs.~\ref{fig:phases05}(b) and (c), the small bubble branch is represented by the red region, while the large bubble branch occupies both blue and gray regions,

\medskip
The solution of Eqs.~(\ref{eq:norm-n1}) and (\ref{eq:def-sigma}) in the range~(\ref{eq:redrange}) provides only a part of the phase diagram  if all parameters remain real (gray region in Fig.~\ref{fig:extension}).
However, it turns out that the solution for $n=-1$ and $\sigma = -0.8$ obtained from Eqs.~(\ref{eq:def-newsigma-nm1}) and (\ref{eq:norm-nm1})  in the same range~(\ref{eq:redrange}) compensates the lacking region via the transformation~(\ref{eq:refsym}) (blue hatched region in Fig.~\ref{fig:extension}). 
These two regions overlap near the $\tilde{\gamma}=0$ curves of each region  (black and blue dashed curves in Fig.~\ref{fig:extension}). 
Both solutions coincide in the overlap region, and hence the entire phase diagram is consistently covered by two regions.

\medskip

The phase diagram for $\sigma = {\rm constant}$ is bounded by the following phases:

\begin{itemize}

\item 
The limit of non-BPS (nonsupersymetric) extremal black hole without bubbles (orange and green solid curves in Fig.~\ref{fig:phases05}), as $\tilde{\gamma} \to \sqrt{2-\tilde{\nu}^2}$ with $\tilde{\nu}$ held constant, where the surface gravity $\kappa$ approaches zero.
 In this limit, the angular momenta correspond to those of the extremal Cveti\v{c}-Youm black holes, maintaining the same charge-to-mass ratio as specified in Eq.~(\ref{eq:qtom-sigma}).
For the limit with smaller $j_\phi$ (orange curves  in Fig.~\ref{fig:phases05}), the angular momenta are given by
\begin{align}\label{eq:bhlim-jj}
j_\psi^{\rm BH} = \frac{j_1+\sigma^3 j_2}{(1+\sigma^2)^{3/2}},\quad
j_\phi^{\rm BH} = \frac{j_2+\sigma^3 j_1}{(1+\sigma^2)^{3/2}},
\end{align}
where
\begin{align}
j_1 &= 1+ j_2\nonum
&=\frac{\left(\sigma ^3+1\right) \left(3
   \left(\sigma ^3-3\right) \tilde{\nu }^2 (\tilde{\nu}^2
   -2)  
   +2   \left(\sqrt{\sigma ^6 \left(-3 \tilde{\nu }^4+6
   \tilde{\nu }^2+1\right)+2 \sigma ^3 \left(3
   \tilde{\nu }^4-6 \tilde{\nu
   }^2+1\right)+1}-\sigma ^3-1\right)\right)}{9
   \tilde{\nu }^2 \left(2-\tilde{\nu }^2\right)},\label{eq:bhlim-j1j2}
\end{align}
and the entropy is given by
\begin{align}
&a_{\cal H}^{\rm BH} = \frac{2 \sqrt{2} \left(\sigma ^3+1\right)^{3/2}}{9 \left(\sigma
   ^2+1\right)^{3/2} \tilde{\nu }^2 \left(2-\tilde{\nu }^2\right)} 
   \biggr[ -9 \sigma ^3 \left(\sigma ^6-5 \sigma ^3+6\right)   \tilde{\nu }^8
   +9 \sigma ^3 \left(\sigma ^6-17 \sigma ^3+30\right) \tilde{\nu }^6\nonum
&\quad   +6 \left(16 \sigma ^9+8   \sigma ^6-83 \sigma ^3-3\right) \tilde{\nu }^4-48 \left(3 \sigma ^9-4 \sigma ^6-8 \sigma ^3-1\right)   \tilde{\nu }^2-32 \left(\sigma ^3+1\right)^3\nonum
&\quad+\left(9 \sigma ^3 \left(\sigma ^3-2\right) \tilde{\nu }^6+\left(-48 \sigma ^6+96 \sigma   ^3+18\right) \tilde{\nu }^4+48 \left(\sigma ^6-3 \sigma ^3-1\right) \tilde{\nu }^2+32 \left(\sigma   ^3+1\right)^2\right)\nonum
&\quad \times  \sqrt{\sigma ^6 \left(-3 \tilde{\nu }^4+6 \tilde{\nu }^2+1\right)+2 \sigma ^3   \left(3 \tilde{\nu }^4-6 \tilde{\nu }^2+1\right)+1}\biggr]^{1/2}. \label{eq:ent-bhlim}
\end{align}

For the limit with larger $j_\phi$ (green curves  in Fig.~\ref{fig:phases05}), the angular momenta are given by
\begin{align}\label{eq:bhlim-jj-2}
j_\psi^{\rm BH} = \frac{j_1-\sigma^3 j_2}{(1+\sigma^2)^{3/2}},\quad
j_\phi^{\rm BH} = -\frac{j_2-\sigma^3 j_1}{(1+\sigma^2)^{3/2}},
\end{align}
where
\begin{align}
j_1 &= 1+ j_2\nonum
&=-\frac{\left(1-\sigma ^3\right) \left(\left(1-\sigma ^3\right) \left(\tilde{\nu }^4-2 \tilde{\nu }^2+2\right)-2 \sqrt{\sigma ^6 \left(\tilde{\nu }^2-1\right)^2-2 \sigma ^3 \left(\tilde{\nu
   }^2-1\right)^2+1}\right)}{\tilde{\nu }^2 \left(2-\tilde{\nu }^2\right)},\label{eq:bhlim-j1j2-2}
\end{align}
and the entropy is given by
\begin{align}
&a_{\cal H}^{\rm BH} =\frac{2 \sqrt{2} \left(1-\sigma ^3\right)^{3/2}}{\left(\sigma ^2+1\right)^{3/2}
   \left(2-\tilde{\nu }^2\right)^2} \biggr[ \sigma ^9 \tilde{\nu}^2 \left(\tilde{\nu }^2-1\right)-3 \sigma ^6 \tilde{\nu
   }^2 \left(\tilde{\nu }^2-1\right)+2 \sigma ^3 \left(\tilde{\nu }^4-\tilde{\nu
   }^2+1\right)-2\nonum
& \quad   +\left(2-\sigma ^3 \left(2-\sigma ^3\right) \tilde{\nu }^2\right)
   \sqrt{\sigma ^6 \left(\tilde{\nu }^2-1\right)^2-2 \sigma ^3 \left(\tilde{\nu
   }^2-1\right)^2+1}\biggr]^{1/2}.
   \end{align}

$\phi_\cD$ and $q_\cD$ take the same form on these curves
\begin{align}
 \phi_\cD^{\rm BH} = \frac{\sqrt{3} \sigma^2}{\sqrt{1+\sigma^2}},\quad q_\cD^{\rm BH} =0.
\end{align}

While, the entropy of the Cveti\v{c}-Youm black hole, in general, is given by~\cite{Cvetic:1996kv}
\begin{align}
a_{\cal H}^{\rm CY} = \frac{\sqrt{2} \left(\sqrt{1-\left(j_1-j_2\right){}^2} \left(1-\sigma ^3\right)+\sqrt{1-\left(j_1+j_2\right){}^2}
   \left(1+\sigma ^3\right)\right)}{\left(1+\sigma ^2\right)^{3/2}},\label{eq:ent-cy}
\end{align}
with $(j_1,j_2)$ in Eq.~(\ref{eq:bhlim-j1j2}) or in Eq.~(\ref{eq:bhlim-j1j2-2}). 
Since the Cveti\v{c}-Youm black hole does not possess the bubble of $S^2$ outside the horizon, both the magnetic flux and magnetic potential are trivially zero.
We note that the limit of the entropy~(\ref{eq:ent-bhlim}) does not coincide with that of of the corresponding Cveti\v{c}-Youm black hole in Eq.~(\ref{eq:ent-cy}).
The similar discontinuity is reported for the black hole limit in the black ring phase~\cite{Elvang:2007hs,Elvang:2004rt}. As speculated in Ref.~\cite{Elvang:2004rt},  this may be because 
the spacetime topology changes at the limit.

\medskip

The two extremal black hole curves intersect at  the limit $\tilde{\nu}\to\sqrt{2}$ ($A$ in  Fig.~\ref{fig:phases05}), 
 where the angular momenta and entropy become
 \begin{align}
( j_{\psi},j_{\phi})\bigr|_A&= \frac{(1,\sigma^3)}{(1+\sigma^2)^{3/2}},\label{eq:orange-p}\\
a_{{\cal H}}\bigr|_A &=0.
\end{align}
The common endpoint $A$ corresponds to the extremal Cveti\v{c}-Youm phase with zero entropy.
Each curve has another endpoint at the limit $\tilde{\nu} \to 1$ ($B$ and $B'$ in Fig.~\ref{fig:phases05}),
where the angular momenta and entropy become
\begin{align}
(j_{\psi},j_{\phi})\bigr|_B&= \frac{(j_{1}+\sigma^3 j_{2}, j_{2}+\sigma^3 j_{1})}{(1+\sigma^2)^{3/2}}\biggr|_B,\\
j_{1}\bigr|_B &= 1+j_{2}\bigr|_B = \biggr\{  \begin{array}{cc} 1 - 9^{-1}(2-\sigma^3)^2 &( 2^{-1/3}<\sigma<1)\\ 1 -\sigma^6 &( -1<\sigma<2^{-1/3}) \end{array},\label{eq:black-p}\\
a_{{\cal H}}\bigr|_B &= 2\sqrt{2}\left(\frac{1+\sigma^3}{1+\sigma^2}\right)^{3/2} \times \quad \biggr\{  \begin{array}{cc} 9^{-1}(2-\sigma^3)\sqrt{2\sigma^3-1} &( 2^{-1/3}<\sigma<1)\\ \sqrt{\sigma^6(1-2\sigma^3)} &  ( -1<\sigma<2^{-1/3}) \end{array}.
\end{align}
and
\begin{align}
(j_{\psi},j_{\phi})\bigr|_{B'}&= \frac{(j_{1}-\sigma^3 j_{2}, -j_{2}+\sigma^3 j_{1})}{(1+\sigma^2)^{3/2}}\biggr|_{B'},\\
j_{1}\bigr|_{B'} &= 1+j_{2}\bigr|_{B'} =1-\sigma^6,\label{eq:black-p}\\
a_{{\cal H}}\bigr|_{B'} &= 2\sqrt{2}\,|\sigma|^3\left(\frac{1-\sigma^3}{1+\sigma^2}\right)^{3/2}. 
\end{align}

\item The limit to a non-BPS (nonsupersymetric), extremal black hole limit with a bubble, i.e., an extremal capped black hole  (black and blue curves in Fig.~\ref{fig:phases05}) as $\tilde{\nu}\to 1$ and $\tilde \gamma\to 1$ with 
\begin{align}\label{eq:extlim}
\tilde{\gamma} = 1 - (\delta/\mu^2) \beta,\quad \tilde{\nu} = 1 + \delta \beta,  \quad (0<\mu<1).
\end{align}
For the limit with smaller $j_\phi$ (with larger $j_\phi$), substituting these expressions into  Eqs.~(\ref{eq:norm-n1}) and  (\ref{eq:def-sigma}) (Eqs.~(\ref{eq:def-newsigma-nm1}) and (\ref{eq:norm-nm1})),  and subsequently taking the limit $\beta \to 0$, we derive these two equations for the parameters $(\sigma,\tilde{b},\delta,\mu)$, solvable for a given $(\sigma,\mu)$ configuration.
These limits intersects with the previous extremal black hole limit at $\mu \to 1$ ($B$ and $B'$ in Fig.~\ref{fig:phases05}) and terminates at $\mu \to 0$ ($C$ and $C'$ in  Fig.~\ref{fig:phases05}).
At the latter points, the spacetime corresponds to the extremal phase with zero entropy.
For instance, at $\sigma=0.8$, the physical parameters at $C$ are numerically determined as follows:
\begin{align}
 (j_{\psi},j_{\phi},\phi_{\cD},q_{\cD})\bigr|_C=(0.482...,0.240...,0.780...,-0.0402...).
\end{align}
At the point  $C'$, the values of $j_\psi,j_\phi$ and $\phi_\cD$ are the same as those at the point $C$, but $q_{\cD}$ are different:
\begin{align}
q_\cD\bigr|_{C'} = -0.0455....
\end{align}

 \item Critical curve (red dot-dashed curves in Fig.~\ref{fig:phases05}), which is defined by the condition
\begin{align}
\partial_{\tilde{\nu}} j_\psi \partial_{\tilde{\gamma}} j_\phi-\partial_{\tilde{\nu}} j_\phi \partial_{\tilde{\gamma}} j_\psi =0,
\end{align}
where $j_\psi$ has a local minimum on a $j_\phi = {\rm constant}$ surface.
This curve forms part of the boundary in the phase diagram only in the $(j_\psi,j_\phi)$ plane, as shown in Fig.~\ref{fig:phases05}(a). 
In the $(\phi_\cD,j_\phi)$ and $(q_\cD,j_\phi)$ planes, it distinguishes between the large and small bubble branches, as illustrated Fig.~\ref{fig:phases05}(b) and (c). 
This curve ends on the extremal black hole curves ($D$ and $D'$ in Fig.~\ref{fig:phases05}). 
The respective physical quantities at these end points are numerically determined as
\begin{align}
 (j_{\psi},j_{\phi},\phi_{\cD},q_{\cD})\bigr|_D&=(0.420...,0.187...,0.866...,0),\\
 \nonum
  (j_{\psi},j_{\phi},\phi_{\cD},q_{\cD})\bigr|_{D'}&=(0.416...,0.303..., 0.866...,0).
 \end{align}

The line of $j_\phi = 0.240\dots$ extending from $C$ to $C'$ in the $(q_\cD,j_\phi)$ plane (gray dashed line in Fig.~\ref{fig:phases05}(c)).
At $\tilde{\nu}\to 1$, the constant curves of $\tilde{\gamma}$ with different values end at different points on this line.
In the $(j_\psi,j_\phi)$ and $(\phi_\cD,j_\phi)$ planes, this curve degenerates to the single point ($C=C'$ in Fig.~\ref{fig:phases05}(a) and (b)), 
 with the value of $\tilde{\gamma}$ determining the direction in which the $\tilde{\gamma}={\rm constant}$ curve approaches this point.

\item

The blue dot-dashed curve in Fig.~\ref{fig:phases05}(c) defined by the condition:
\begin{align}
 \partial_{\tilde{\nu}} j_\phi \partial_{\tilde{\gamma}} q_\cD
 - \partial_{\tilde{\nu}} q_\cD \partial_{\tilde{\gamma}} j_\phi=0,
\end{align}
where $q_\cD$ has a local maximum on the $j_\phi={\rm constant}$ phase (Fig.~\ref{fig:qd-profile}). 
As in Fig.~\ref{fig:phases05}(c), this curve forms a part of the boundary in the $(q_\cD,j_\phi)$ plane, 
stretching from the point $E$ on the line of $j_\phi = 0.240\dots$ to the point $F$ at which it intersects with the curve of the extremal capped black hole limit,  as illustrated in Fig.~\ref{fig:phases05}(c), where
\begin{align}
& (j_{\psi},j_{\phi},\phi_{\cD},q_{\cD})\bigr|_{E}=( 0.482..., 0.240..., 0.780...,-0.0420...),\\
\nonum
& (j_{\psi},j_{\phi},\phi_{\cD},q_{\cD})\bigr|_{F}=( 0.481..., 0.239..., 0.781...,-0.0398...).
\end{align}
\begin{figure}
\includegraphics[width=7cm]{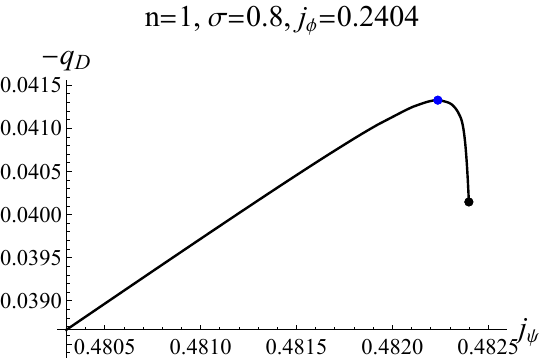}
\caption{$q_{\cD}$ on the $j_\phi=0.2404$ phase near the extremal capped black hole limit (black dot) for $n=1$ and $\sigma =0.8$. The local maximum (blue dot) corresponds to the blue dot-dashed curve in Fig.~\ref{fig:phases05}(c).
\label{fig:qd-profile}}
\end{figure}

\end{itemize}

\begin{figure}[t]
\includegraphics[width=7cm]{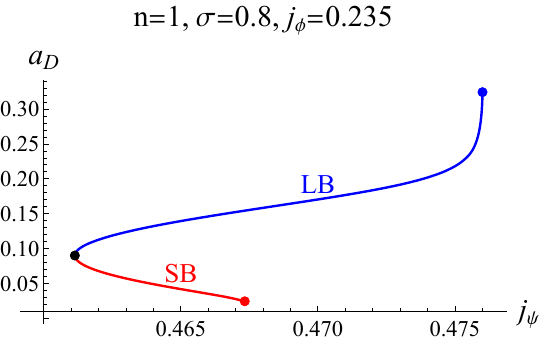}
\caption{Bubble area of large (blue) and small (red) bubble branches for $\sigma=0.8$ and $j_\phi=0.235$. \label{fig:bubblesize} }
\end{figure}

\subsection{Nonuniqueness of spherical black holes: large and small bubble branches}

In our previous works~\cite{Suzuki:2023nqf,Suzuki:2024phv}, 
we have demonstrated that the capped black hole with three independent parameters can carry conserved charges--mass, two angular momenta, and electric charge-- identical to those of the Cveti\v{c}-Youm black hole. 
This reveals a level of non-uniqueness among spherical black holes, extending beyond the known non-uniqueness attributed to the existence of different topological black objects such as black holes and black rings. 
Furthermore, the present capped black hole with four independent parameters surprisingly possesses further two branches: one with a large bubble and another with a small bubble, yet both admitting the same conserved charges.

\medskip

As in Fig.~\ref{fig:phases05}, the capped black hole solution displays discrete non-uniqueness relative to the charges $(M, J_\psi, J_\phi, Q)$, 
and supports two distinct solutions: large bubble branch and small bubble branch, differentiated by the relative sizes of their bubble areas. 
Specifically, the large bubble branch features a greater bubble area compared to its counterpart. 
This difference is quantified using the dimensionless bubble area defined as:
\begin{align}
a_\cD := \frac{A_\cD}{2\pi r_M^2} ,\quad 
\end{align}
where $A_\cD$ is defined in Eq.~(\ref{eq:def-bubble-area}).
In Fig.~\ref{fig:bubblesize}, we compare the bubble area of the large and small branches for a given $(\sigma,j_\psi,j_\phi)$.

\begin{figure}[t]
\includegraphics[width=7cm]{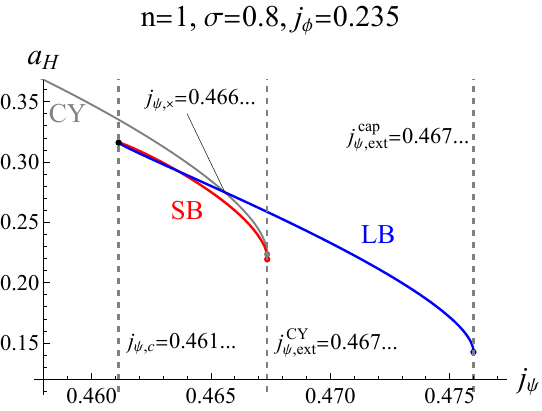}
\caption{Entropy of large and small bubble branches for $n=1$, $\sigma=0.8$ and $j_\phi=0.235$ displayed by the blue and red curves, respectively. 
The entropy of the Cveti\v{c}-Youm black hole with the same charges is also shown by the gray curve. \label{fig:entropy} }
\end{figure}

\medskip
Figure~\ref{fig:entropy} compares the entropy of the large, small bubble branches, and the Cveti\v{c}-Youm black hole 
of the same mass-to-charge ratio given by Eq.~(\ref{eq:qtom-sigma}) with $\sigma =0.8$ having same angular momenta $(j_\psi,j_\phi=0.235)$,
where we label the values of $j_\psi$ at the cross section of the extremal capped black hole limit, extremal black hole limit, and the critical curve as $j_{\psi,{\rm ext}}^{\rm cap}$, $j_{\psi,{\rm ext}}^{\rm CY}$, and $j_{\psi,c}$, respectively.
We observe that three distinct solutions are present for $j_{\psi,c} < j_{\psi} < j_{\psi, {\rm ext}}^{\rm CY}$, whereas only the large bubble branch exists for $j_{\psi, {\rm ext}}^{\rm CY} < j_{\psi} < j_{\psi, {\rm ext}}^{\rm cap}$. 
The area where three solutions are possible corresponds to the purple region  in Fig.~\ref{fig:phases05}(a).
In Fig.~\ref{fig:phases05}(b) and (c), the small bubble branch is depicted in red, and the corresponding large bubble branch is depicted in blue.
These regions are separated from the  region admitting only the large bubble branch (gray region in Fig.~\ref{fig:phases05}) by the curve $j_\psi=j_\psi^{\rm BH}(\sigma,j_\phi)$ (orange dashed curves in Fig.~\ref{fig:phases05}(b) and (c)) where $j_\psi^{\rm BH}$ is defined by Eq.~(\ref{eq:bhlim-jj}).

\medskip

Moreover, the  value $j_{\psi,\times}$ can be identified as the phase where the entropies of the Cveti\v{c}-Youm black hole and the large bubble branch are equal. 
For $j_{\psi,c} < j_\psi< j_{\psi,\times}$ , the Cveti\v{c}-Youm black hole is thermodynamically favored due to its higher entropy, while for $j_{\psi,\times} < j_\psi < j_{\psi,{\rm ext}}^{\rm CY}$, the large bubble branch has the greater entropy and is thus preferred. 
The small bubble branch occasionally exhibits higher entropy than the large bubble branch near the critical curve, but it remains thermodynamically irrelevant as it has lower entropy compared to the Cveti\v{c}-Youm black hole.

\subsection{Relation to the three-parameter solution in Ref.~\cite{Suzuki:2023nqf,Suzuki:2024phv}}

\begin{figure}[t]
\includegraphics[width=5.5cm]{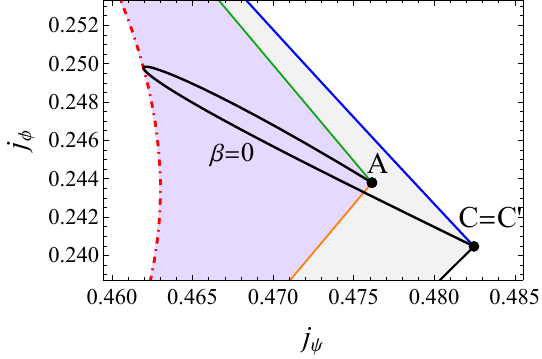}
\includegraphics[width=5.5cm]{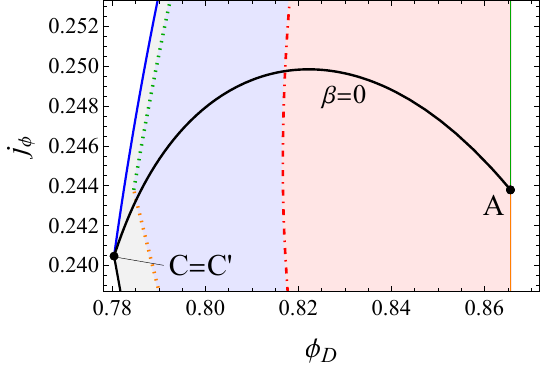}
\includegraphics[width=5.5cm]{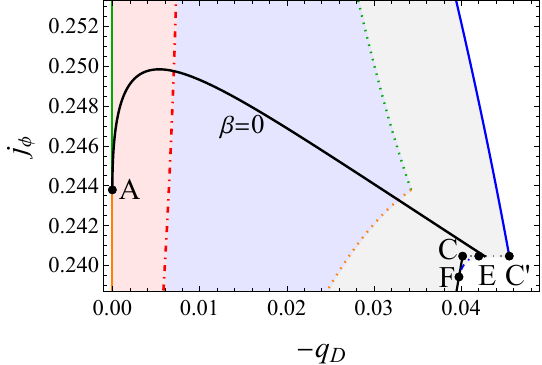}
\caption{The phase of the three parameter capped black hole solution (black curve) embedded in the four parameter phase space for $n=1$ and $\sigma=0.8$ presented in Fig.~\ref{fig:phases05}.
\label{fig:3param}}\end{figure}

\medskip
Finally, we discuss the connection between the new four-parameter solution and the previous three-parameter solution, as detailed in Refs.\cite{Suzuki:2023nqf,Suzuki:2024phv}. 
This three-parameter solution is derived from the four-parameter solution by simply setting $\beta=0$. 
As shown in Fig.~\ref{fig:3param}, the $\beta=0$ phase is represented by a curve starting from $A$ with $\tilde{\nu}=\sqrt{2}$ and ending at $C=C'$ with $\tilde{\nu}=1$ in the $(j_\psi,j_\phi)$ and $(\phi_\cD, j_\phi)$ planes.
Whereas, in the $(q_\cD, j_\phi)$ plane, the end point with $\tilde{\nu}=1$ appears on the $j_\phi=0.240\ldots$ line between $C$ and $C'$.
Because the $\beta=0$ phase does not exhibit multiple branches in the $(j_\psi,j_\phi)$-plane,
the non-uniqueness of the capped black hole, discussed in the previous subsection, cannot be confirmed within three parameter-parameter solutions.

%%%%%
\section{Summary}\label{sec:sum}.

%In this paper, we have constructed a capped black hole, which is a non-BPS solution of an asymptotically flat, stationary, bi-axisymmetric, spherical black hole with the DOC of topology $[\mathbb{R}^4 \# \mathbb{CP}^2] \setminus {\mathbb B}^4$ on a timeslice within five-dimensional minimal supergravity. 
In this paper, we have constructed the  new non-BPS solution of the capped black hole  in the bosonic sector of five-dimensional minimal supergravity, which describes an asymptotically flat, stationary, bi-axisymmetric, charged spherical black hole with the DOC of the topology of $[\mathbb{R}^4 \# \mathbb{CP}^2] \setminus {\mathbb B}^4$ on a timeslice.
Using the Harrison transformation, we have previously constructed the three-parameter family solution of the capped black hole~\cite{Suzuki:2023nqf,Suzuki:2024phv}. By combining the Ehlers and Harrison transformations, we could have generalized it to the four-parameter family solution, allowing all conserved charges to be independent.
We have found that the new solution displays discrete non-uniqueness with respect to the conserved
 charges, allowing two distinct branches for the same mass, angular momenta, and electric charge. These branches, called the large bubble and small bubble branches, differ in the size of the bubble adjacent to the horizon. 
We have also found that the Cveti\v{c}-Youm black hole exists for the conserved charges with which two capped black hole branches exist. Whereas, with the angular momenta beyond the extremal limit of the Cveti\v{c}-Youm black hole, there exists only the large bubble branch.
Our analysis have shown that, among the three, the most thermodynamically preferred solution is the large bubble branch if the bubble size is sufficiently large, but otherwise the Cveti\v{c}-Youm black hole.
While the small bubble branch always has smaller entropy than the Cveti\v{c}-Youm black hole,
 it can have larger entropy than the large bubble branch, at least for $\sigma=0.8$ near the critical curve.

\medskip

We have found that, as long as keeping all parameters real, the $Q/M={\rm constant}$ phase in the $n=1$ case is not complete, but it also requires the solution obtained from the $Q/M={\rm constant}$ phase in the $n=-1$ case via the transformation $(t,\psi)\to(-t,-\psi)$.
The complete phase space may be expressed by using only the $n=1$ solution if one can extend the parameter region to include certain complex parameters,  as preformed in Ref.~\cite{Suzuki:2024vzq}. 
This extension deserves further investigation in our future work.
It may also be an  interesting issue to construct non-BPS black hole solutions with the DOC of other non-trivial topology,  $[{\mathbb R}^4 \# n {\mathbb C}{\mathbb P}^2 \# m (S^2\times S^2)]\setminus {\mathbb B}^4$ with $(n,m)\not =(1,0)$  on a timeslice. 
This extension will be the focus of our subsequent studies.

\section*{Acknowledgement}
R.S. was supported by JSPS KAKENHI Grant Number JP24K07028. 
S.T. was supported by JSPS KAKENHI Grant Number 21K03560.

\appendix

\section{Coefficients $f_i$ and $g_i$}\label{sec:fi-gi}
The explicit forms of the coefficients $f_i$ and $g_i$ are shown below. 

\subsection{$f_1,\dots,f_{19}$}
Every $f_i$ is a quadratic function of $\beta$. $f_i$ for $16\leq i\leq 19$ appears only in the expressions of $g_i$.
\begin{align}
 f_1& = 1+2( a- b) \beta +\frac{\beta ^2}{\gamma^2 -\nu^2 } \biggr[(a-b)^2 \gamma ^2-(1-a)^2 \gamma  (1+2 b+(3+2 b) \nu )\nonum
&\quad\quad+\nu  \left((1-a)^2 (3+2 b)+\left(1-2 a+2 \left(1-a+a^2\right) b-b^2\right) \nu \right)\biggr],\\
 f_2 &= 1+2(1- a b) \beta +\frac{\beta ^2}{\gamma^2 -\nu^2 } \biggr[ \left((2-a) a-2 \left(1-a+a^2\right) b-(1- 2 a) b^2\right) \gamma ^2\nonum
&\quad \quad +(1-a)^2 \gamma  (1+2 b+(3+2 b) \nu   )-\nu  \left((1-a)^2 (3+2 b)+(1-b) (1+(1-2 a) b) \nu \right)\biggr],\\
 f_3 &= 1+2( a- b) \beta +\frac{\beta ^2}{\gamma^2 -\nu^2 } \biggr[ (a-b)^2 \gamma ^2-2 (1-a)^2 (1+a)+(1-a)^2 (1-2 a+2 b) \nu \nonum
&\quad\quad+ \left(1-2 a+2 \left(1-a+a^2\right)   b-b^2\right) \nu ^2+(1-a)^2 \gamma  (1+2 a-2 b-(1-2 a+2 b) \nu )\biggr],\\
 f_4 &= 1- \frac{2 \beta}{\gamma -\nu }  \left(a^2-1-a (a-b) \gamma +\nu(1 -a b) \right)+\frac{\beta ^2}{\gamma^2 -\nu^2 } \biggr[ 2 (1-a)^2 (1+a)-(1-2 a) (a-b)^2 \gamma   ^2\nonum
&\quad \quad -(1-a) (1-5 a+4 b) \nu -\left(1-2 a b-b^2+2 a b^2\right) \nu ^2-(1-a) \gamma  \left(1-a-4 a^2+4 a b-(1 -5 a  +4 b) \nu   \right)\biggr],\\
 f_5& = 1   - \frac{2 \beta }{c_2} \left(2 a   (-1+\gamma ) \nu +a b^2 (\gamma -\nu ) (1+\nu )+b \left(-1+a^2-a^2 \gamma -\left(2+a^2\right) (-1+\gamma ) \nu +\nu ^2\right)\right)\nonum
& \quad +\frac{\beta ^2}{(\gamma^2 -\nu^2 )  c_2} \biggr[ b \gamma  \left(2 (1-a)^2 (1+a)+(1-a) \left(-1+a+4 a^2-4 a b\right) \gamma +(-1+2 a) (-a+b)^2 \gamma ^2\right)\nonum
& \quad +\nu \biggr\{   \left(2 \left(-2+a+6 a^2-4 a^3\right)+\left(4-20 a+17 a^2-6   a^3\right) b+4 \left(2-3 a+2 a^2\right) b^2+(1-2 a) b^3\right) \gamma ^2\nonum
& \quad \quad 
+(-a+b)^2 (-4+2 a-b+2 a b) \gamma ^3-2   (1-a)^2 (1+a) (2+b)+2 (1-a)^2 \left(4+5 a+(-1+3 a) b-2 b^2\right) \gamma\biggr\} \nonum
& \quad    +  \nu ^2 \biggr\{  (1-a)   \left(2 \left(2-3 a+a^2\right)+\left(3-9 a+2 a^2\right) b +4 b^2\right)\nonum
& \quad \quad +\left(4 (a-2) (1-a)^2+\left(-10+26 a-21 a^2+4 a^3\right) b-2   \left(4-7 a+2 a^2\right) b^2+(1-2 a) b^3\right) \gamma\nonum
& \quad \quad +\left(2 (2-a) (1-a)^2+\left(5-14 a+10 a^2-2 a^3\right) b+\left(4-6 a+4 a^2\right)   b^2+(1-2 a) b^3\right) \gamma ^2\biggr\}\nonum
& \quad   - \left(2 a-\left(6-6 a+5 a^2\right) b+4 a b^2+(1-2 a) b^3\right) (1-\gamma ) \nu ^3+(1-b) b (1+(1-2 a) b) \nu ^4 \biggr],\\
 f_6&= 1-\frac{2 \beta}{(\gamma -\nu ) (1+\nu ) c_1}  \biggr[ a^2-1+\left(2-2 a+a^2+b-2 a b\right) \nu
 -(1-b) (1-2 a+b)   \nu ^2-(1-b) b \nu ^3\nonum
 &\quad\quad +(a-b)^2 \gamma ^2 (1+\nu )+\gamma  \left(1-2 a^2-b+2 a b-2   \left(1-a+a^2-2 a b+b^2\right) \nu +(1-b) (1-2 a+2 b) \nu   ^2\right)\biggr]\nonum
&\quad   +\frac{\beta ^2 }{(\gamma^2 -\nu^2 ) (1+\nu ) c_1}\biggr[
2 (1-a)^2   (1+a)-(1-a) \left(2-5 a-a^2+4 b\right) \nu \nonum
&\quad\quad+\left(-2-2 a+6 a^2-3 a^3+\left(3-4  a+3 a^2\right) b+(2-3 a) b^2\right) \nu ^2+\left(2-3 a+3 a^2 b-3 a b^2+b^3\right)
   \nu ^3+(1-b)^2 b \nu ^4\nonum
&\quad\quad +(a-b) \gamma ^2 \left(4 a^2-1-2a+(2-3 a) b+\left((2-a)^2-2 a b+b^2\right) \nu -\left(3 (1-a)^2+(2-a)b-b^2\right) \nu ^2\right)\nonum
&\quad\quad     -(a-b)^3 \gamma ^3 (1+\nu )+ (1-a) \gamma \left(a-2+5 a^2-4 a   b+\left(2-9 a-a^2+8 b\right) \nu\right) \nonum
&\quad\quad+\left(2+5 a-12 a^2+6   a^3-\left(8-12 a+7 a^2\right) b+3 a b^2-b^3\right) \gamma \nu ^2-\left(2-3 a+3 a^2 b-3 a   b^2+b^3\right) \gamma \nu ^3\biggr], 
\end{align}
\begin{align}
 f_7&=  1-\frac{2 \beta }{b (\gamma -\nu ) (1+\nu )} \biggr[(b-1) b \gamma +\left(2-2 a+b-b^2\right) (1-\gamma ) \nu
   +(1-b) b \nu ^2 \biggr]\nonum
&   +\frac{\beta ^2}{b   (\gamma^2 -\nu^2 ) (1+\nu )}
 \biggr[b \gamma    \left((1-a)^2+\left((2-a) a-2 b+b^2\right) \gamma \right)-(1-b)^2 b \nu ^3\nonum
 &\quad
 +\left(-(1-a)^2   (4+3 b)+4 (1-a) (1-2 a+(2-a) b) \gamma  +\left(4 (1-a) a-\left(4-6   a+a^2\right) b-2 b^2+b^3\right) \gamma ^2\right) \nu \nonum
 &\quad
   +\left(4 (-1+a)+(2-3 a) a b+2   b^2-b^3+(1-a) (4-b-3 a b) \gamma \right) \nu ^2\biggr],\\
 f_8&= 1+\frac{2 \beta}{(a-b) (\gamma -\nu )}\biggr[1-a^2+(a-b)^2 \gamma -\left(1-2 a b+b^2\right) \nu\biggr]\nonum
&   -\frac{\beta ^2}{(a-b)   \left(\gamma ^2-\nu ^2\right)}\biggr[
2(-1+ a+ a^2- a^3)-a \gamma( 1  + 2 a - 3 a^2) -(a^3-b^3) \gamma ^2-b^2 \gamma    \left(2+2 a^2+a (-4+3 \gamma )\right)\nonum
&\quad
+b \left(-2 (1-a)^2   (1+a)+\left(1+4 a-7 a^2+2 a^3\right) \gamma +3 a^2 \gamma ^2\right)\nonum
&\quad
- \nu (1-a) (\gamma -1) \left((a-5) a+\left(5-3 a+2   a^2\right) b+2 (1-a) b^2\right)  \nonum
&\quad
-\left(-2+a+\left(-1+6 a-2   a^2\right) b+\left(-2+a-2 a^2\right) b^2+b^3\right) \nu ^2\biggr],\\
 f_9&= 1+2 (a-b) \beta 
+\frac{\beta ^2}{\left(\gamma ^2-\nu ^2\right) c_1}\biggr[
-2 (1-a)^2 (1+a)-(a-b)^3 \gamma ^3+(1-a)^2  (4+a+2 b) \nu-(1-b) b (1-2 a+b) \nu ^3\nonum
& -\left(2-5 a+4 a^2+\left(1-2 a-a^2\right) b+a b^2\right) \nu   ^2
-(a-b) \gamma ^2 \left(1-a (2-b)-\left((1-a)^2-a   b+b^2\right) \nu \right)\nonum
&  -\gamma  \left((1-a)^2 (-2-3 a+2 b)+2 \left(2-3
   a+a^3\right) \nu +\left(-2+5 a-4 a^2+\left(-2+4 a+a^2\right) b-3 a b^2+b^3\right)
   \nu ^2\right)\biggr],\\
 f_{10}&=\frac{4 (1-a)  (1-\gamma ) d_1}{\gamma^2 -\nu^2 }\beta  (1+(a-b) \beta ),\\
 f_{11}&=1-\frac{2 \beta  (-1+a-a \gamma +b (-1+a+\gamma )+(-1+a b) \nu )}{\gamma +\nu   }\nonum
& +\frac{\beta ^2}{\gamma^2 -\nu^2 } \biggr[(a-b)^2 \gamma ^2
+ \nu  \left((1-a) \left(a-3+2  b^2\right)-(1-b) (1+b-2 a b) \nu \right)+(1-a) \gamma  \left(1+a-2  b^2+\left(3-a-2 b^2\right) \nu \right)\biggr],\\
f_{12}&=f_1+\frac{c_3 \left(f_1-f_3\right)}{(a+1) b (1-\gamma ) (\nu +1)},\\
f_{13}&=f_7+\frac{\left(b (\gamma-\nu^2) -(2+b) (1-\gamma ) \nu \right) f_{10}}{2 b (1+\nu ) d_1},
\end{align}
\begin{align}
 f_{14}&=1-\frac{2 \beta }{d_2} \biggr[ b \left(-1+a^2+\gamma -2 a^2 \gamma -(1-2 a) b \gamma   +(a-b)^2 \gamma ^2\right)\nonum
 &\quad+\left(-2+\left(2+a^2\right) b+(1-2 a) b^2-2   \left(-1+b+a^2 b-2 a b^2+b^3\right) \gamma +b (-a+b)^2 \gamma ^2\right) \nu
\nonum
&\quad   +\left(2-b-2 a b^2+b^3+\left(-2+b+(1+2 a) b^2-2 b^3\right) \gamma \right) \nu
   ^2+(-1+b) b^2 \nu ^3\biggr]\nonum
&   +\frac{\beta ^2}{\left(\gamma ^2-\nu ^2\right) d_2} \biggr[
b \gamma  \left(2 (1-a)^2   (1+a)+(1-a) \left(-2+a+5 a^2-4 a b\right) \gamma \right. \nonum
&\quad\quad\left. +\left(a \left(-1-2 a+4   a^2\right)+\left(1+4 a-7 a^2\right) b+(-2+3 a) b^2\right) \gamma ^2-(a-b)^3   \gamma ^3\right) \nonum
&\quad +\left(-2 (1-a)^2 (1+a) (2+b)+2 (1-a)^2 \left(5+6 a+4 a b-2   b^2\right) \gamma \right.\nonum
&\quad \quad\left. +\left(-6+4 a+16 a^2-12 a^3+\left(2-22 a+26 a^2-11 a^3\right)
   b+\left(9-16 a+11 a^2\right) b^2+(2-3 a) b^3\right) \gamma ^2\right.\nonum
&\quad\quad\left. -2 \left((3-2 a)
   a^2-3 a \left(2-2 a+a^2\right) b+\left(3-4 a+5 a^2\right) b^2-3 a b^3+b^4\right)
   \gamma ^3+b (-a+b)^3 \gamma ^4\right) \nu\nonum
&\quad  +\left((1-a) \left(2-6 a+4   a^2+\left(2-9 a+3 a^2\right) b+4 b^2\right)\right.\nonum
&\quad \quad
+\left(6 (1-a)^2 (-1+2   a)+\left(-12+37 a-34 a^2+8 a^3\right) b+\left(-13+20 a-5 a^2\right) b^2+(2-3 a)
   b^3\right) \gamma \nonum
&\quad \quad+\left(4-24 a+30 a^2-12 a^3+\left(18-39 a+32 a^2-7 a^3\right)
   b+2 \left(5-10 a+4 a^2\right) b^2\right) \gamma ^2\nonum
&\quad \quad\left.+\left(2 a \left(4-5 a+2   a^2\right)+\left(-8+13 a-10 a^2+2 a^3\right) b+\left(-3+4 a-3 a^2\right) b^2+(2+3
   a) b^3-2 b^4\right) \gamma ^3\right) \nu ^2\nonum
&\quad +\left(2 (a-2) a-\left(-10+14 a-10   a^2+a^3\right) b+\left(3-8 a+a^2\right) b^2+(-2+3 a) b^3\right.\nonum
&\quad \quad\left.
+2 \left(-1+4 a-2   a^2+\left(-8+14 a-10 a^2+a^3\right) b-\left(3-8 a+a^2\right) b^2-3 a
   b^3+b^4\right) \gamma \right.\nonum
&\quad \quad\left. +\left(2 (-1+a)^2-\left(-6+14 a-10 a^2+a^3\right)
   b+\left(5-8 a+a^2\right) b^2+(-2+3 a) b^3\right) \gamma ^2\right) \nu   ^3-(1-b)^2 b^2 \nu ^5\nonum
&\quad +\left(2-(2+a) b+\left(2-4 a+a^2\right) b^2+3 a b^3-b^4+\left(-2+(2+a)
   b-\left(1-4 a+a^2\right) b^2-(2+3 a) b^3+2 b^4\right) \gamma \right) \nu   ^4  \biggr],\\
f_{15}&=1-\frac{2 \beta }{(\gamma -\nu ) \left(c_3-(1-a b) (1-\gamma ) (1+\nu )\right)}\biggr[
1-a^2-\left(2+a^2+a \left(-2-2 b+b^2\right)\right) \nu \nonum
&+\left(1-2 (1-a) b-(1+a)   b^2+b^3\right) \nu ^2-(1-b) b^2 \nu ^3-(a-b) \gamma ^2 \left(a-b+b^2+\left(-2+a-b+b^2\right) \nu
   \right)\nonum
&+\gamma  \left(-1+2 a^2-2 a b+a b^2+2 \left((1-a)^2+b-2 a b+(1+a) b^2-b^3\right) \nu +\left(-1+2   (1-a) b+(2+a) b^2-2 b^3\right) \nu ^2\right)\biggr] \nonum
& +\frac{\beta ^2}{(\gamma^2 -\nu^2 ) \left(c_3-(1-a b) (1-\gamma )   (1+\nu )\right)}
 \biggr[ -2 (1-a)^2 (1+a) (1+b)\nonum
 &\quad-(1-a) \left(1+5 a-2 a^2+\left(-4+3 a-3 a^2\right) b-2 (1-a)   b^2\right) \nu \nonum
 &\quad +\left(4-6 a+3 a^2-\left(-2+8 a-4 a^2+a^3\right) b+\left(2-2 a+3 a^2\right) b^2-a b^3\right)
   \nu ^2 \nonum
 &\quad+\left(-1+(4-3 a) b+\left(-1+2 a+a^2\right) b^2-(2+a) b^3+b^4\right) \nu ^3+(-1+b)^2 b^2 \nu
   ^4 \nonum
 &\quad+(a-b)^2 \gamma ^3 \left(-1+2 a+(-2+a) b+b^2+\left(-3+2 a+(-2+a) b+b^2\right) \nu \right) \nonum
 &\quad+\gamma ^2
   \left(-1+2 a+4 a^2-6 a^3-\left(2+5 a-14 a^2+4 a^3\right) b+\left(4-10 a+3 a^2\right) b^2+a b^3\right.\nonum
   &\quad \quad \left. +\left(-2-4
   a+13 a^2-6 a^3+\left(8-22 a+18 a^2-5 a^3\right) b+\left(7-14 a+5 a^2\right) b^2+(2+a) b^3-b^4\right) \nu\right.\nonum
   &\quad \quad \left. 
   +\left(3 (-1+a)^2-(-2+a) (-1+a)^2 b+\left(1-4 a+2 a^2\right) b^2+2 b^3-b^4\right) \nu ^2\right) \nonum
 &\quad+\gamma 
   \left((-1+a) \left(-3+a+6 a^2+\left(-4-5 a+5 a^2\right) b-2 (-1+a) b^2\right)\right.\nonum
   &\quad \quad \left. +(-1+a) \left(-3-11 a+6
   a^2+\left(12-11 a+7 a^2\right) b-6 (-1+a) b^2\right) \nu\right.\nonum
   &\quad \quad \left.  -\left(7-12 a+6 a^2+\left(4-13 a+8 a^2-2 a^3\right)
   b+\left(5-6 a+5 a^2\right) b^2-(2+a) b^3+b^4\right) \nu ^2\right.\nonum
   &\quad \quad \left. +\left(1+(-4+3 a) b-\left(-1+2 a+a^2\right)
   b^2+(2+a) b^3-b^4\right) \nu ^3\right) \biggr],        
   \end{align}
    \begin{align}
f_{16} &= 1-\frac{2 \beta  \left(-1+a^2+(a-b)^2 \gamma ^2+(1-2 a+b) \nu +(1-b) b \nu ^2-\gamma  \left(-1+2 a^2+b-2 a
   b+(1-2 a+b) \nu \right)\right)}{(\gamma +\nu ) c_1}\nonum
&   +\frac{\beta ^2}{\left(\gamma ^2-\nu ^2\right) c_1}\biggr[ 2 (1-a)^2 (1+a)-(a-b)^3 \gamma ^3-(1-a) \left(4+a^2-a (1+4 b)\right) \nu \nonum
&\quad+\left(2-3 a-\left(1-4 a+a^2\right) b+(-2+a) b^2\right) \nu   ^2+(1-b)^2 b \nu ^3\nonum
&\quad+(a-b) \gamma ^2 \left(-1-2 a+4 a^2+(2-3 a) b+\left((-1+a)^2+(2-3 a) b+b^2\right) \nu
   \right)\nonum
&\quad+\gamma  \left((1-a) \left(-2+a+5 a^2-4 a b\right)+2 (1-a) \left(2-a+a^2-4 a b+2 b^2\right) \nu\right)
\nonum
&\quad   -\left(2-3 a+(4-a) a b-(4-a) b^2+b^3\right)\gamma \nu ^2\biggr],\\    
 f_{17}& =1+\frac{2 \beta  (1-a+(a-b) \gamma +(1-b) \nu )}{\gamma +\nu }\nonum
&+\frac{\beta ^2 \left((a-b)^2 \gamma ^2
+ \nu    \left((1-a) (1-3 a+2 b)-(1-b)^2 \nu \right)+(1-a) \gamma  (1+a-2 b-\nu +3 a \nu -2 b \nu   )\right)}{\gamma^2 -\nu^2},       \\
f_{18}& =1+\frac{2 \beta  \left(1-a^2+a (a-b) \gamma +(2-a) (a-b) (1-\gamma ) \nu -(1-a b) \nu ^2\right)}{(1-\nu )
   (\gamma +\nu )}\nonum
&   + \frac{\beta ^2}{(\gamma^2 -\nu^2 ) (1-\nu )} \biggr[ 2 (1-a)^2 (1+a)-(1-a) \left(3+2 a^2-a (1+4 b)\right) \nu +\left((a-2)   a+2 (3-2 a) a b-(3-2 a) b^2\right) \nu ^2\nonum
&\quad   +(1-b) (1+(1-2 a) b) \nu ^3+(a-b)^2 \gamma ^2 (-1+2 a+(-3+2 a)   \nu )\nonum
&\quad+(1-a) \gamma  \left(-1+a+4 a^2-4 a b+\left(2-2 a+4 a^2-8 a b+4 b^2\right) \nu +\left(-1+a-4 a b+4   b^2\right) \nu ^2\right) \biggr],   \\
f_{19}&=1+\frac{2 \beta  (1-a+(a-b) \gamma -(1-b)\nu  )}{\gamma -\nu }+\frac{\beta ^2 \left(1-a^2-2 b (1-a (1-\gamma ))+a^2 \gamma +b^2 \gamma -(1-b)^2 \nu \right)}{\gamma -\nu   }.
\end{align}

%%%%%%%%%%
% g_i
%%%%%%%%%%

\subsection{$g_1,\dots,g_{13}$ and $\tilde{g}_1$}
For the metric functions, we have
 \begin{align}
g_1 & =f_{13} f_{14}-\frac{c_1 c_3 \left(f_{13} f_{14}-f_{16} f_{19}\right)}{2 (a-b) (1-\gamma )^2 \nu }+\frac{(1-\nu )
   c_3 f_3 f_{10}}{2 (a-b) (1-\gamma ) (1+\nu ) d_1},
\label{eq:def-g1}\\
g_2&=\Delta+\frac{c_3 \left(\Delta-f_3 f_4\right)}{d_3}, \label{eq:def-g2}\\
g_3 &=f_{14}f_{18}-\frac{d_1 \left(f_{14} f_{18}-f_2 f_{17}\right)}{(1-\gamma ) (1+\nu ) \left(a   c_1-1+\nu \right)}-\frac{b \nu  (\gamma -\nu ) f_{10} f_{12}}{(1-\gamma ) (1-\nu^2 ) \left(a   c_1-1+\nu \right)},\\
g_4 &=\Delta+\frac{1}{d_4}\biggr[-(1-\gamma )^2 (1-\nu )^2 \nu  d_1 \left(\Delta -f_1^2\right)
-\frac{8 \nu  d_1 \left(d_1-(1+a) (1-\gamma )   (1+\nu ) c_1\right) \left(\Delta -f_1 f_2\right)}{1+a}\nonum
&+2 (1-\gamma ) (1-\nu ) \nu  \left(-2-3 \nu +3 \nu   ^2\right) d_1 \left(\Delta -f_2^2\right)-\frac{1}{2} (1-\nu ) (3-\nu ) d_2^2
   \left(\Delta-f_{14}^2\right)\nonum
&-\frac{2 (\gamma -\nu ) (1-\nu )^3 c_3 \left(c_3+ b(1+a)  (1-\gamma ) (1+\nu   )\right) \left(\Delta -f_3 f_4\right)}{1+a}+2 (1-\gamma ) (1-\nu )^2 \nu  c_2^2   \left(\Delta -f_5^2\right)\nonum
&+\frac{1}{2} (\gamma -\nu ) (1+\nu ) \left(3-2 \nu +\nu ^2\right) c_1^2 c_3
   \left(\Delta -f_6^2\right)
   +(1-\nu ) \nu ^2 \left(c_3-2 (1-\gamma ) \nu \right) d_1 \left(\Delta -f_7^2\right)\nonum
   &+2
   (a-b)^2 (1-\gamma )^3 (1-\nu ) (\gamma -\nu ) \nu ^3 \left(\Delta -f_8^2\right)+\frac{(\gamma -\nu ) (-1+\nu )^2
   \nu ^2 (\gamma +\nu ) c_3 f_{10}^2}{(1+\nu ) d_1}\biggr], \label{eq:def-g4}\\
g_{5} & =\Delta -\frac{d_1 \left(\Delta -f_1 f_2\right)}{a d_1+\left(1-a^2\right) (1-\gamma ) (1+\nu ) c_1},\\   
g_6 &=f_8^2+\frac{2 (1-\gamma )^2 (1-\nu )^2 \nu  d_1 \left(f_1^2-f_8^2\right)}{(1+\nu ) d_5}+\frac{(1-\gamma )^2   (1-\nu )^4 c_3 \left(f_3^2-f_8^2\right)}{(1+\nu ) d_5},\\
g_7 &= g_8 +\frac{c_3 f_{10}^2 (\nu -1)^2 \nu  \left(\gamma ^2-\nu ^2\right)}{d_1 d_6 (\nu +1)},\\
g_8 &=f_{14}^2-\frac{2 (1-\gamma ) (1-\nu )^2 \nu  d_1 \left(f_{14}^2-f_2^2\right)}{d_6}+\frac{(\gamma -\nu )
   (1-\nu )^2 (1+\nu ) c_1^2 c_3 \left(f_{14}^2-f_6^2\right)}{2 d_6 \nu },\\
g_9&=f_8 f_{18} +\frac{(2+\nu ) f_1 f_{10}}{(a-b) (1-\gamma ) (1-\nu^2 )},\\
g_{10}&=f_{14} f_{18}-\frac{b f_{10} f_{12} (\nu +1) (\gamma -\nu )}{2 d_2 (\nu -1)},\\
g_{11}&=f_{16}  f_{15}-\frac{(1-\gamma ) \left(4+\nu -\nu ^2\right) f_1 f_{10}}{(1+\nu ) \left(2 (a-b) (1-\gamma )^2   (1-\nu ) \nu -b (1+\nu ) d_1\right)}-\frac{(a-b) (1-\gamma )^2 (1-\nu )^2 \left(f_{18} f_8-f_{16} f_{15}\right)}{2 (a-b) (1-\gamma )^2 (1-\nu ) \nu -b (1+\nu ) d_1}.
\end{align}
Additionally, we also define
\begin{align}
\tilde{g}_1 &=  g_{1}+\frac{\nu c_3 f_3 f_{10} }{(1-\gamma ) d_1 (\nu +1) (a-b)}\nonum
&= f_{13} f_{14}+ \frac{c_3 f_3 f_{10}}{2 (a-b) (1-\gamma ) d_1}-\frac{c_1 c_3\left(f_{13} f_{14}-f_{16}   f_{19}\right)}{2 (a-b) (1-\gamma )^2 \nu },\\
g_{12} &= f_{5}f_{13}+\frac{c_3(f_4 f_{17}-f_{5}f_{13})}{2\nu (1-\gamma)(1-ab)},\\
  g_{13} &= f_8 f_{18}-\frac{f_1 f_{10}}{2 (1-\gamma ) (1-\nu ) (b-a)}.
\end{align}

\end{document}